# Star Formation and Molecular Gas Diagnostics with Mid- and Far-infrared Emission

C. M. Whitcomb[1], K. Sandstrom[2], A. Leroy[3], and J.-D. T. Smith[1]
[1] Ritter Astrophysical Research Center, Department of Physics and Astronomy, University of Toledo, Toledo, OH 43606, USA; coryw777@gmail.com
[2] Center for Astrophysics and Space Sciences, Department of Physics, University of California San Diego, 9500 Gilman Drive, La Jolla, CA 92093, USA
[3] Department of Astronomy, The Ohio State University, 140 West 18th Avenue, Columbus, Ohio 43210, USA

## Abstract

With the start of JWST observations, mid-infrared (MIR) emission features from polycyclic aromatic hydrocarbons (PAHs), $H_2$ rotational lines, fine structure lines from ions, and dust continuum will be widely available tracers of gas and star formation rate (SFR) in galaxies at various redshifts. Many of these tracers originate from dust and gas illuminated by UV photons from massive stars, so they generally trace both SFR and gas to varying degrees. We investigate how MIR spectral features from 5–35 $\mu$m and photometry from 3.4–250 $\mu$m correlate with SFR and molecular gas. In general, we find MIR emission features (i.e., PAHs and $H_2$ rotational lines) trace both CO and SFR better than CO and SFR trace one another. $H_2$ lines and PAH features correlate best with CO. Fine structure lines from ions correlate best with SFR. The [S III] lines at 18.7 and 33.5 $\mu$m, in particular, have a very tight correlation with SFR, and we use them to calibrate new single-parameter MIR tracers of SFR that have negligible metallicity dependence in our sample. The 17 $\mu$m/7.7 $\mu$m PAH feature ratio increases as a function of CO emission which may be evidence of PAH growth or neutralization in molecular gas. The degree to which dust continuum emission traces SFR or CO varies as a function of wavelength, with continuum between 20 and 70 $\mu$m better tracing SFR, while longer wavelengths better trace CO.

## 1. Introduction

The molecular gas content of a galaxy and the rate at which it forms stars are crucial properties that govern its evolution. A large amount of effort has been dedicated to characterizing the scaling relations that describe star formation (SF) by comparing measurements of SF and gas for integrated galaxies and resolved regions of galaxies (e.g., the Kennicutt–Schmidt (K-S) relation, Schmidt 1959; Kennicutt 1998; Saintonge et al. 2011; de los Reyes & Kennicutt 2019; Kennicutt & De 2021). Observations show that on large scales in galaxies there is a strong correlation between gas and SF (Wong & Blitz 2002; Bigiel et al. 2008; Leroy et al. 2008). Due to the effects of cloud evolution and SF feedback, these relationships break down on small scales, revealing spatially distinct regions of molecular and ionized gas (Schruba et al. 2010; Kruijssen & Longmore 2014; Schinnerer et al. 2019; Chevance et al. 2020; Kim et al. 2022). These studies make use of a variety of tracers for SF and gas, and there are many complications in calibrating each (e.g., Kennicutt & Evans 2012; Calzetti 2013; Tacconi et al. 2020; Saintonge & Catinella 2022; Belfiore et al. 2023), including dust attenuation for SF tracers and systematic effects of metallicity, excitation, and more for gas tracers.

Mid-infrared (MIR) tracers of $H_2$ and SFR could potentially be ideal as they suffer minimal dust attenuation. In addition, high-sensitivity, high-angular resolution MIR observations of galaxies at a wide range of redshifts are now possible with JWST. Many studies have investigated MIR emission lines, polycyclic aromatic hydrocarbon (PAH) vibrational bands, and dust continuum as tracers of SF and molecular gas (e.g., Calzetti et al. 2005; Wu et al. 2005; Alonso-Herrero et al. 2006; Murphy et al. 2011; Jarrett et al. 2013; Maragkoudakis et al. 2018; Gao et al. 2019; Lai et al. 2020; Chown et al. 2021; Gao et al. 2022). These MIR tracers, however, will generally have sensitivity to both dust content, and by extension gas content, as well as the strength of the UV radiation field illuminating the interstellar medium (which is related to SF).

For example, $H_2$ rotational lines originate in molecular gas, but are powered by UV radiation through far-UV (FUV) pumping and gas heating leading to collisional excitation (Kaufman et al. 2006; Roussel et al. 2007). Likewise, PAHs seem to be largely absent from the ionized gas in H II regions (Povich et al. 2007; Chastenet et al. 2019), but emit brightly in the intensely UV-irradiated neutral and molecular gas surrounding them (i.e., dense photodissociation regions (PDRs; Tielens 2008). Some MIR fine structure lines originate in both ionized and neutral gas (i.e., lines of $Si^+$, which have an ionization potential <13.6 eV), leading to sensitivity both to the ionizing and FUV photons from young massive stars, and to the properties of neutral gas in PDRs. Because of the joint relationships with gas and SF for many MIR tracers, assessing the SFR–$H_2$ scaling relationships becomes more complex.

Even with a clear spatial distinction between tracers of SF and $H_2$, the fact that tracers respond on different timescales to SF can cause additional decorrelation in the K-S relationship (e.g., Leroy et al. 2012). Ionized gas tracers respond on the timescale of O star lifetimes, <10 Myr. Tracers sensitive to FUV photons (e.g., $H_2$, PAH bands, hot dust continuum) can still be excited by B stars and have lifetimes up to ∼100 Myr. As a result, even two observables that both trace SF can decorrelate due to the fact that we catch star-forming regions at different times after SF episodes (e.g., Lee et al. 2009). For some *hybrid* SF tracers, which combine UV/optical emission with MIR dust continuum to correct for dust attenuation, a concern is whether the UV/optical and dust emission trace SF on different timescales (e.g Crocker et al. 2013; Mallory et al. 2022). This is closely related to the issue of *cirrus* contamination of IR-based SF indicators, where dust emission powered by longer-lived stars contaminates SF tracers (Leroy et al. 2012; Boquien et al. 2016). For all of these reasons, studies quantifying the relationship between tracers of gas and SF are critical.

Early JWST results are providing insights into these questions already. Recent work by Leroy et al. (2023) investigated the use of MIR PAH emission as a tracer of gas and SF at high resolution with JWST photometric observations. They found that PAH emission traces both gas and SF, since PAHs are generally well mixed with the gas and respond approximately linearly to the intensity of the UV radiation field when they are stochastically heated. These observations agree well with expectations from dust spectral energy distribution modeling, which suggest a significant fraction of the IR emission arises from dust heated by the average interstellar radiation field (Draine et al. 2007; Aniano et al. 2020).

In the following, we address similar questions for multiple MIR SF and gas tracers using spectroscopic constraints from archival Spitzer observations. To study the degree to which individual MIR emission features and IR photometric bands can be used to independently trace molecular gas or SFR, we measure the emission features in MIR spectra from the Spitzer Infrared Nearby Galaxies Survey (SINGS; Kennicutt & Armus 2003). Our approach is to investigate the joint SF–$H_2$ correlations for individual MIR emission features and dust continuum emission.

We attempt to differentiate tracers that follow molecular gas content from those that trace SF by comparing correlation coefficients with cold gas and SF respectively. To first order we expect correlations between all MIR SF and molecular gas tracers due to the observed K-S relation between $\Sigma_{\rm SFR}$ and $\Sigma_{\rm H2}$ (e.g., Bigiel et al. 2008; Leroy et al. 2008; Kennicutt & Evans 2012; Leroy et al. 2013). These correlations will be enhanced by the fact that some features trace both SF and $H_2$ simultaneously. If we look at smaller spatial scales (sub-kiloparsec), the breakdown in the K-S relationship (Schruba et al. 2010; Kennicutt & Evans 2012; Kruijssen & Longmore 2014; Schinnerer et al. 2019) provides some insights to differentiate features that trace SF from those that trace $H_2$. At scales that resolve H II regions, diffuse gas, and molecular clouds, this distinction will be maximized, though the decorrelation due to the timescale of each tracer will still be important. The H II regions we study here are not highly resolved, but the majority have sub-kiloparsec resolution, where there is an increase in the scatter of the K-S relation due to the scale breakdown.

This paper is organized as follows: Section 1.1 provides context on the tracers of SF and gas we explore in this paper. In Section 2 we describe our sample selection, data extraction criteria, and our fitting and correlation methodology. Section 3 describes our correlation results with CO and SFR. Section 4 describes the physical implications of our results and how they compare with those from previous studies. Section 5 outlines our most significant conclusions.

### *1.1. Background and Motivation*

In the following we test a variety of MIR features and their joint correlations with SF and gas; here we briefly review some key points about them and their origins.

#### *1.1.1. Tracing SFR*

In extragalactic observations, the star formation rate (SFR) is often calculated based on ultraviolet (UV) emission from young stars or emission from the ionized gas surrounding them. Both can be difficult to measure directly due to dust attenuation, so hybrid UV-infrared or H$\alpha$-infrared tracers are frequently used (see review by Kennicutt & Evans 2012). Such tracers are generally empirically calibrated. Energy conserving models (e.g., MAGPHYS; CIGALE, da Cunha et al. 2008; Boquien et al. 2019) can be used instead to account for dust attenuation with a more physically realistic approach. However, these models require sufficient coverage of the UV to IR spectral energy distribution.

Ionized gas can also be observed at longer wavelengths where dust attenuation is negligible. One such ionized gas tracer is the thermal component of radio continuum (i.e., free–free emission) at around 33 GHz (Condon 1992; Murphy et al. 2011; Linden et al. 2020). But emission at this frequency is intrinsically faint and can be difficult to detect even in nearby galaxies.

In the MIR there are several strong fine structure lines that trace ionized gas. For example, Ne $\rightarrow$ $Ne^+$ and $S^+ \rightarrow S^{2+}$ have ionization potentials greater than that of hydrogen, at 21.6 and 23.3 eV, respectively, so their emission lines originate from ionized gas in H II regions. MIR fine structure lines are important for cooling in H II regions and some are known to correlate well with recent SF activity (Ho & Keto 2007; Zhuang et al. 2019; Whitcomb et al. 2020). However, SF tracers using individual fine structure lines often have scatter that depends on the metallicity and hardness of the local radiation field. In this study, we use as our reference tracer a combination of the 15.6 $\mu$m [Ne III] and 12.8 $\mu$m [Ne II] lines with a small metallicity correction (see Section 2.6).

The primary motivation in using this fine structure line SFR tracer is that previous work calibrated it against free–free radio continuum emission in the SINGS sample (Whitcomb et al. 2020). Since free–free radio continuum is not subject to most of the systematic effects that complicate SFR tracers based on ionized gas, it serves as a *gold standard*. The attenuation is also negligible in the MIR, so the reference tracer used here is independent of dust emission. Because our aim is to distinguish between correlations with SFR and cold gas (and therefore dust) it is critical to use an SFR tracer that is minimally influenced by dust, both in attenuation and by any *hybrid SFR*-type correction. The neon-based SFR tracer is ideal for this since it was calibrated with free–free emission and has no dust-corrected component. Finally, the 15.6 $\mu$m [Ne III] and 12.8 $\mu$m [Ne II] lines are derived from the same MIR spectra as the dust and other features in this study. This removes concerns related to calibration differences and allows us to use all SINGS spectroscopy regions in our study.

Many studies have used dust emission in the infrared to trace reprocessed UV photons from obscured, young stars. In particular, total infrared (TIR) emission is widely used as an SFR tracer (Kennicutt 1998; Galametz et al. 2013). This tracer is dependent on the assumption that the TIR emission is dominated by emission from dust grains heated by UV photons from young stars, and that a constant fraction of the UV photons is reprocessed to the IR. But older stellar populations can also be a significant source of the radiation that drives the dust emission (e.g., Boquien et al. 2016). This complicates the use of dust emission as a tracer of recent SF.

The dust continuum emission as measured by Spitzer at 24 $\mu$m is another well-established tracer of obscured SF at galactic as well as sub-kiloparsec scales (Calzetti et al. 2007; Kennicutt et al. 2009; Leroy et al. 2012). Previous studies have found that emission at 24 $\mu$m traces SF on $\sim$100 Myr timescales (Calzetti et al. 2007).

The WISE 12 $\mu$m band captured the complex of PAH features between 11 and 12 $\mu$m and has been used to calibrate both molecular gas and SF tracers (Donoso et al. 2012; Jarrett et al. 2013; Gao et al. 2019; Chown et al. 2021; Gao et al. 2022). The IRAC 8 $\mu$m band is usually dominated by the strongest emission feature at 7.7 $\mu$m from PAHs, which are likewise supposed to trace UV photons produced primarily by O stars in regions of ongoing SF. However, in addition to significant contributions to PAH excitation from older stellar populations (e.g., Crocker et al. 2013; Smercina et al. 2018), these PAH-based tracers are also found to be highly dependent on metallicity (Engelbracht et al. 2005; Calzetti et al. 2007; Draine et al. 2007; Li 2020; Whitcomb et al. 2020). In addition, unlike the larger dust grains that dominate the IR continuum emission, the PAH molecules are susceptible to ionizing radiation of sufficient energy that can erode them or inhibit their formation (Micelotta et al. 2010a, 2010b). In this study, we focus on H II regions where the fraction of PAH emission originating from the star-forming region is expected to be significantly greater than the fraction of emission from the diffuse interstellar medium (Draine et al. 2007).

#### *1.1.2. Tracing $H_2$*

Cold $H_2$ content is typically inferred using low-$J$ rotational lines of $^{12}C^{16}O$. This requires a "CO-to-$H_2$" ($\alpha_{CO}$) conversion factor, which can vary significantly within and between galaxies (Bolatto et al. 2013). The gas content can also be traced with dust. By modeling the far-infrared (FIR) continuum emission, the dust mass can be inferred and, using a dust-to-gas ratio, can be converted to gas mass (Draine 2003; Leroy et al. 2011; Sandstrom et al. 2013; Galliano et al. 2018; Chiang et al. 2021).

The MIR rotational lines of $H_2$ itself arise from the warmer fraction of molecular gas and they are relatively bright in MIR spectra. These emission lines originate predominantly from warm regions (100–1000 K), which implies a potentially complicated relationship to a tracer of cold gas (20–100 K) such as CO emission (Roussel et al. 2007). Previous work has shown that modeling the temperature distribution of the $H_2$ rotational lines can accurately recover the total $H_2$ content from the $\sim$15% of the $H_2$ in this warm phase (Togi & Smith 2016). An additional complication is that $H_2$ rotational excitation in star-forming regions can be driven by UV heating, either directly through a cascade from electronically excited states through higher-$J$ states (UV pumping) or indirectly through photoelectric heating of the gas. Therefore, $H_2$ rotational lines likely depend on both SF and gas content. The $H_2$ can also be rotationally excited by strong turbulence or shocks (Ingalls et al. 2011; Appleton et al. 2017; Smercina et al. 2018).

Emission from PAHs is likely connected to both SFR and cold gas content to varying degrees. PAHs exist in neutral and molecular gas and only emit when excited by UV/optical photons. However, the amount of PAHs also increases at higher column densities of cold gas, which will likewise result in more PAH emission. At the moderate optical depths of a typical dense PDR the local radiation field is composed of photons that are very effective at exciting the PAHs but not so strong as to destroy them. In this warm portion of the dense PDR, photoelectrons from PAHs and dust grains are the dominant source of gas heating at UV optical depths $\tau \sim 1$–5 where temperatures are $\sim$100–1000 K (Wolfire et al. 2022). PAHs are also found to emit from diffuse regions where they are still exposed to infrequent UV photons, or plentiful optical photons, from the interstellar radiation field of nearby stellar populations (Draine et al. 2007; Bendo et al. 2008; Ingalls et al. 2011). One of the primary goals of this study is to determine if MIR, and in particular PAH, emission is a better tracer of the cold gas content or of SFR in normal star-forming regions.

## 2. Data and Analysis Methods

### *2.1. IRS Spectroscopy*

The SINGS program observed nuclear and extranuclear H II regions in 75 nearby galaxies with the Short-Low (SL2 5.2–7.7 and SL1 7.4–14.7 $\mu$m), Short-High (SH 9.9–19.4 $\mu$m), Long-Low (LL 14–38 $\mu$m), and Long-High (LH 19.1–37.1 $\mu$m) modules of the InfraRed Spectrograph (IRS) on the Spitzer Space Telescope (Kennicutt & Armus 2003; Houck et al. 2004; Smith et al. 2007; Dale et al. 2009). We use data from SINGS Data Release 5.[4] The LL spatial coverage of extranuclear regions in SINGS is sparse compared to SH and LH, so we make use of LH for the long wavelength coverage. Our sample is therefore composed of the overlapping areas between the SL, SH, and LH data for each nuclear and extranuclear target. We note that this data set is focused on optically bright HII regions in local spiral galaxies. This biases our sample toward more massive and evolved H II regions.

A similar data set was defined in Roussel et al. (2007) in order to study the $H_2$ rotational lines. Our selection is nearly the same as that in Roussel et al. (2007) except we use the Data Release 5 of the SINGS data and the apertures are defined independently. The resulting spectral range includes all the major PAH features (except 3.3 $\mu$m, which contributes $\sim$2% of the total PAH power; Lai et al. 2020) and the 12.8 $\mu$m [Ne II] and 15.6 $\mu$m [Ne III] lines that are needed for our SFR tracer (see Section 2.6). We extract spectra in all IRS orders from the same apertures. The full data set of extraction apertures consists of 154 polygonal apertures with numbers of vertices ranging from 4–7. A table with the vertices of each region and all data used in the following analysis is available in machine-readable form. A description of the contents of this table is shown in Appendix A.1.

### *2.2. MIR and FIR Photometry*

We use Spitzer, Wide-field Infrared Survey Explorer (WISE), and Herschel IR photometric observations from 3.4–250 $\mu$m from SINGS, the "$z = 0$ Multiwavelength Galaxy Synthesis" (z0MGS; Leroy et al. 2019), and "Key Insights on

[4] https://irsa.ipac.caltech.edu/data/SPITZER/SINGS/doc/sings_fifth_delivery_v2.pdf

Nearby Galaxies: a Far-Infrared Survey with Herschel" (KINGFISH; Kennicutt et al. 2011). We use the SINGS photometric data from the four InfraRed Array Camera (IRAC) bands at 3.6, 4.5, 5.8, and 8 $\mu$m, and the 24 $\mu$m Multiband Imaging Photometer for Spitzer (MIPS) band. We did not apply extended source correction factors to this data since they do not affect our correlation analysis. The z0MGS data includes the four WISE bands at 3.4, 4.6, 12, and 22 $\mu$m. We also include FIR KINGFISH data from the three Photodetector Array Camera and Spectrometer (PACS) photometric bands at 70, 100, and 160 $\mu$m, as well as the Spectral and Photometric Imaging Receiver (SPIRE) 250 $\mu$m photometric band (Kennicutt et al. 2011).

We convolve all photometric data to 15″ to match the resolution of our CO data, except SPIRE 250 $\mu$m which has 18″ resolution (see Section 2.4). We perform a simple background subtraction on each photometric image using the median value of the image outside the $R_{25}$ isophotal radius of the galaxy with a 3$\sigma$ mask to filter out background galaxies and foreground stars. We multiply each photometric band by its characteristic frequency in order to match the units of integrated intensities for spectral emission features ($10^{-7}$ W m$^{-2}$ sr$^{-1}$) to facilitate direct comparisons between photometry and spectroscopy. We also use the photometric bands longer than 20 $\mu$m to calculate TIR using the calibration adapted from Galametz et al. (2013; see Section 2.6). We assume a constant 10% of the measured surface brightness in each aperture for all photometric bands. We find our results do not vary significantly if this assumption is doubled.

### 2.3. CO Mapping

The molecular gas data used in this work consists of CO $J = (2-1)$ measurements from the "Physics at High-Angular resolution in Nearby GalaxieS"—Atacama Large Millimeter Array (PHANGS-ALMA) and the "Heterodyne Receiver Array CO-Line Extragalactic Survey" (HERACLES) from the compilation in Leroy et al. (2022). These CO maps have been convolved to a Gaussian point-spread function (PSF) with FWHM of 15″. We leave the maps at this resolution in order to match the resolution of FIR maps from KINGFISH. Of the 75 galaxies in the SINGS sample, we have a CO $J = (2-1)$ map for 41. Each map has a corresponding error map from which we extract the same IRS overlap apertures to determine the uncertainty on the corresponding CO emission. Of the 154 regions in our IRS spectroscopy sample, 104 have a CO measurement, and 89 of these have detections of CO with a signal-to-noise ratio (S/N) >3.

We use CO emission as our reference molecular gas tracer and we do not convert to $H_2$ gas masses with a CO-to-$H_2$ conversion factor. Since the CO-to-$H_2$ conversion is approximately the same for most of our regions, this choice will not affect the correlation coefficients. We explore the expected effect of translating CO emission to $M_{H2}$ in Section 4.2.

### 2.4. Convolution and Aperture Corrections

We extract a single average spectrum from all Spitzer-IRS orders in our apertures based on the overlap of SL1, SL2, SH, and LH spatial coverage. The size of these apertures is comparable to the angular resolution of the CO data (e.g., ∼15″ in diameter). To ensure matched extractions for the photometry, we convolve each IRAC, WISE, MIPS, and PACS image to a Gaussian PSF with FWHM of 15″ matched to the CO maps using convolution kernels from Aniano et al. (2011). We note SPIRE 250 $\mu$m has a slightly lower resolution (∼18″) that cannot be matched to 15″ but is close enough that we do not expect major issues, so we leave it at the native PSF.

For the spectroscopic data, the coverage of the spectral maps is small, so we cannot convolve to matched 15″ resolution without introducing artifacts. Our apertures are larger than the PSF of the IRS data at all wavelengths (Pereira-Santaella et al. 2010), which should minimize any PSF effects, so we proceed to extract and average the spectra in these regions without PSF matching. To check that PSF effects are minimal and no aperture corrections are needed for the IRS spectral extraction, we compare the native resolution Spitzer photometric maps (which have similar PSFs to the spectroscopy) to 15″ resolution convolved images at the IRAC 5.8 and 8 $\mu$m bands and the MIPS 24 $\mu$m band. We extract photometry in our apertures from the native and convolved images and determine the native-to-convolved ratio for each region. The native-to-convolved ratio is found to vary little with wavelength: $1.1 \pm 0.1$ for each photometric band. We then interpolate the ratio as a function of wavelength from 5.8–24 $\mu$m to estimate the effect of using the native resolution spectra in comparison with the 15″ photometric maps. We find our final results are not altered if we use this aperture correction on the IRS data, so we proceed without applying any correction.

### 2.5. Additional Measurements

We determine metallicities for our apertures using the central values and gradients from the Kobulnicky & Kewley (2004; KK04) calibration found in Moustakas et al. (2010). We find no significant difference in our observed trends in Section 3 if the Pilyugin & Thuan (2005; PT05) calibrated metallicities are used instead of the KK04 values. For galaxies with no calculated gradient we use the characteristic value, and for those with no characteristic we use the luminosity–metallicity (L-Z) value. These metallicities are used in combination with the 12.8 $\mu$m [Ne II] and 15.6 $\mu$m [Ne III] lines to determine $\Sigma_{SFR}$ according to the prescription from Whitcomb et al. (2020; see Section 2.6), which was also calibrated with KK04 metallicities from Moustakas et al. (2010).

We use nuclear classifications from Murphy et al. (2018) and Moustakas et al. (2010) to identify apertures that include an active galactic nucleus (AGN). These regions are excluded from our results. For simplicity, we consider galaxies indicated as "SF/AGN" (see Table 1) as AGN and exclude their centers as well. From the 154 regions in our sample, only 10 do not have significant detections of either of the two neon lines required for our SFR calibration (see Section 2.6). Of the remaining 144, 32 are centers of galaxies containing an AGN. We remove these AGN from our sample, leaving a total of 112 SF regions and of these there are 75 that are covered by CO mapping and 67 that have CO detections at S/N >3. We use the full set of 112 regions for SFR correlations and CO correlations only use the subset of 67 regions. Including the weaker detections as upper limits does not significantly constrain the slope and intercept of our power-law fits, so we exclude them from the analysis.

We also determine the deprojected physical area of our apertures in order to investigate how spatial scale affects

**Table 1**
Galaxy Information

| Galaxy | R.A. (J2000) | Decl. (J2000) | Dist. (Mpc) | $i$ (deg) | P.A. (deg) | $R_{25}$ (arcmin) | Characteristic (12+log[O/H]) | Gradient (dex/$R_{25}$) | Central (12+log[O/H]) | Nuc. Type |
|---|---|---|---|---|---|---|---|---|---|---|
| DDO 053 | 08 34 06.50 | 66 10 48.00 | 3.56 | 30 | 121 | 0.77 | 8.00 ± 0.09 | ⋯ | ⋯ | SF |
| DDO 165 | 13 06 24.92 | 67 42 24.95 | 4.57 | 58 | 90 | 1.73 | 8.04 ± 0.07 | ⋯ | ⋯ | SF |
| Ho II | 08 19 04.35 | 70 43 18.03 | 3.05 | 38 | 16 | 3.97 | 8.13 ± 0.11 | ⋯ | ⋯ | SF |
| Ho IX | 09 57 31.97 | 69 02 45.46 | 3.70 | 36 | 0 | 1.26 | 8.98 ± 0.05 | ⋯ | ⋯ | SF |
| IC 2574 | 10 28 23.50 | 68 24 43.54 | 3.79 | 68 | 50 | 6.59 | 8.24 ± 0.11 | ⋯ | ⋯ | SF |
| IC 4710 | 18 28 37.96 | −66 58 56.10 | 9.00 | 40 | 4 | 1.82 | 8.74[a] | ⋯ | ⋯ | SF |
| Mrk 33 | 10 32 31.88 | 54 24 03.75 | 22.9 | 21 | 150 | 0.50 | 8.87 ± 0.02 | ⋯ | ⋯ | SF |
| Tol 89 | 14 01 33.50 | −33 05 32.00 | 16.7 | 54 | 172 | 1.41 | 8.69 ± 0.06 | ⋯ | ⋯ | SF |
| NGC 0024 | 00 09 56.34 | −24 57 49.57 | 7.30 | 82 | 46 | 2.88 | 8.93 ± 0.11 | ⋯ | ⋯ | SF |
| NGC 0337 | 00 59 50.04 | −07 34 40.86 | 19.3 | 52 | 130 | 1.44 | 8.84 ± 0.05 | ⋯ | ⋯ | SF |
| NGC 0628 | 01 36 41.80 | 15 47 00.45 | 7.20 | 24 | 25 | 5.24 | ⋯ | −0.57 ± 0.04 | 9.19 ± 0.02 | SF |
| NGC 0855 | 02 14 03.60 | 27 52 36.85 | 9.73 | 71 | 67 | 1.32 | 8.80 ± 0.09 | ⋯ | ⋯ | SF |
| NGC 0925 | 02 27 16.90 | 33 34 44.41 | 9.12 | 57 | 102 | 5.24 | ⋯ | −0.42 ± 0.02 | 8.91 ± 0.01 | SF |
| NGC 1097 | 02 46 19.10 | −30 16 30.17 | 14.2 | 48 | 130 | 4.67 | ⋯ | −0.29 ± 0.09 | 9.17 ± 0.01 | AGN |
| NGC 1377 | 03 36 39.06 | −20 54 07.24 | 24.6 | 62 | 92 | 0.89 | 8.89[a] | ⋯ | ⋯ | SF |
| NGC 1482 | 03 54 38.91 | −20 30 08.41 | 22.6 | 57 | 103 | 1.23 | 8.95 ± 0.08 | ⋯ | ⋯ | SF |
| NGC 1566 | 04 20 00.39 | −54 56 16.11 | 20.4 | 38 | 60 | 4.16 | 9.34[a] | ⋯ | ⋯ | AGN |
| NGC 1705 | 04 54 13.50 | −53 21 39.69 | 5.10 | 43 | 50 | 0.95 | 8.28 ± 0.04 | ⋯ | ⋯ | SF |
| NGC 2403 | 07 36 51.42 | 65 36 08.71 | 3.22 | 57 | 128 | 10.9 | ⋯ | −0.26 ± 0.03 | 8.89 ± 0.01 | SF |
| NGC 2915 | 09 26 11.81 | −76 37 35.33 | 3.78 | 61 | 130 | 0.95 | 8.28 ± 0.08 | ⋯ | ⋯ | SF |
| NGC 2976 | 09 47 15.51 | 67 54 58.55 | 3.55 | 65 | 143 | 2.94 | 8.98 ± 0.03 | ⋯ | ⋯ | SF |
| NGC 3031 | 09 55 33.16 | 69 03 54.99 | 3.55 | 60 | 157 | 13.5 | ⋯ | −0.45 ± 0.07 | 9.11 ± 0.02 | AGN |
| NGC 3049 | 09 54 49.61 | 09 16 17.05 | 19.2 | 49 | 25 | 1.09 | 9.10 ± 0.01 | ⋯ | ⋯ | SF |
| NGC 3184 | 10 18 16.94 | 41 25 27.63 | 11.7 | 21 | 135 | 3.71 | ⋯ | −0.52 ± 0.05 | 9.30 ± 0.02 | SF |
| NGC 3198 | 10 19 54.92 | 45 32 59.01 | 14.1 | 70 | 35 | 4.26 | ⋯ | −0.66 ± 0.11 | 9.10 ± 0.03 | SF |
| NGC 3265 | 10 31 06.78 | 28 47 47.73 | 19.6 | 40 | 73 | 0.64 | 8.99 ± 0.06 | ⋯ | ⋯ | SF |
| NGC 3351 | 10 43 57.70 | 11 42 13.26 | 9.33 | 48 | 13 | 3.71 | ⋯ | −0.15 ± 0.03 | 9.24 ± 0.01 | SF |
| NGC 3521 | 11 05 48.75 | −00 02 05.13 | 11.2 | 64 | 163 | 5.48 | ⋯ | −0.69 ± 0.20 | 9.20 ± 0.03 | AGN |
| NGC 3627 | 11 20 15.01 | 12 59 29.77 | 9.38 | 65 | 173 | 4.56 | 8.99 ± 0.10 | ⋯ | ⋯ | AGN |
| NGC 3773 | 11 38 12.93 | 12 06 43.28 | 12.4 | 32 | 165 | 0.59 | 8.92 ± 0.03 | ⋯ | ⋯ | SF |
| NGC 3938 | 11 52 49.40 | 44 07 14.78 | 17.9 | 24 | 29 | 2.69 | 9.06[a] | ⋯ | ⋯ | AGN |
| NGC 4236 | 12 16 42.11 | 69 27 45.39 | 4.45 | 74 | 162 | 10.9 | 8.74[a] | ⋯ | ⋯ | SF |
| NGC 4254 | 12 18 49.62 | 14 25 00.45 | 14.4 | 30 | 24 | 2.69 | ⋯ | −0.42 ± 0.06 | 9.26 ± 0.02 | SF/AGN |
| NGC 4321 | 12 22 54.88 | 15 49 20.10 | 14.3 | 32 | 30 | 3.71 | ⋯ | −0.35 ± 0.13 | 9.29 ± 0.04 | AGN |
| NGC 4559 | 12 35 57.68 | 27 57 34.86 | 6.98 | 68 | 150 | 5.36 | ⋯ | −0.36 ± 0.07 | 8.92 ± 0.02 | SF |
| NGC 4625 | 12 41 52.72 | 41 16 25.93 | 9.30 | 30 | 28 | 1.09 | 9.05 ± 0.07 | ⋯ | ⋯ | SF |
| NGC 4631 | 12 42 07.84 | 32 32 33.06 | 7.62 | 89 | 86 | 7.74 | 8.75 ± 0.09 | ⋯ | ⋯ | SF |
| NGC 4736 | 12 50 53.10 | 41 07 13.09 | 4.66 | 36 | 105 | 5.61 | ⋯ | −0.11 ± 0.15 | 9.04 ± 0.01 | AGN |
| NGC 5055 | 13 15 49.30 | 42 01 45.40 | 7.94 | 56 | 105 | 6.30 | ⋯ | −0.54 ± 0.18 | 9.30 ± 0.04 | AGN |
| NGC 5194 | 13 29 52.69 | 47 11 42.54 | 7.62 | 20 | 163 | 5.61 | ⋯ | −0.50 ± 0.05 | 9.33 ± 0.01 | AGN |
| NGC 5408 | 14 03 21.13 | −41 22 37.51 | 4.81 | 62 | 62 | 0.81 | 8.23 ± 0.06 | ⋯ | ⋯ | SF |
| NGC 5713 | 14 40 11.51 | −00 17 20.41 | 21.4 | 27 | 10 | 1.38 | 9.03 ± 0.03 | ⋯ | ⋯ | SF |
| NGC 6822 | 19 44 56.98 | −14 48 01.23 | 0.47 | 50 | 122 | 7.74 | 8.67 ± 0.10 | ⋯ | ⋯ | SF |
| NGC 6946 | 20 34 52.60 | 60 09 12.66 | 6.80 | 32 | 53 | 5.74 | ⋯ | −0.28 ± 0.10 | 9.13 ± 0.04 | SF |
| NGC 7552 | 23 16 10.81 | −42 35 03.26 | 21.0 | 38 | 1 | 1.69 | 9.16 ± 0.01 | ⋯ | ⋯ | SF |
| NGC 7793 | 23 57 49.82 | −32 35 27.87 | 3.91 | 48 | 98 | 4.67 | ⋯ | −0.36 ± 0.07 | 8.98 ± 0.02 | SF |

**Notes.** Metallicity, gradient, and $R_{25}$: Moustakas et al. (2010), inclinations, position angles, and distances: Murphy et al. 2018, nuclear designations: Moustakas et al. 2010 cross-checked and supplemented by Murphy et al. (2018)
[a] L-Z determined values where no characteristic or $R = 0$ value is given.

correlations between MIR, CO, and SF. This requires the galaxy inclinations, position angles, and distances listed in Table 1. The median area of the 112 regions is about 0.4 kpc$^2$ and likewise for the 67 of these regions with CO data. We compare the behavior of correlations involving the half of regions greater than 0.4 kpc$^2$ to those with smaller areas in Section 3.1. Since the SINGS apertures were placed based on peaks of H$\alpha$ emission (Kennicutt & Armus 2003), we expect that in at least half the regions with area <0.4 kpc$^2$ the $\Sigma_{\rm SFR}$ is resolved distinctly from the CO and the two are spatially uncorrelated (Schinnerer et al. 2019).

### 2.6. Derived Quantities

From the above-described data, we derive TIR for each of our apertures using the calibration presented in Galametz et al. (2013):

$$\begin{aligned}\mathrm{TIR} = 2.013 \times (\nu S_{24}) + 0.508 \times (\nu S_{70}) \\ + 0.393 \times (\nu S_{100}) + 0.599 \times (\nu S_{160}) \\ + 0.680 \times (\nu S_{250}),\end{aligned} \qquad (1)$$

where the constants are taken from their Table 3, $S_\lambda$ is the surface brightness at $\lambda$ microns in MJy sr$^{-1}$, and $\nu$ is the frequency corresponding to the characteristic wavelength of each band. We use this calibration for all galaxies for simplicity instead of applying galaxy-specific calibrations where they are available.

We also derive SFR for each of our apertures using the calibration presented in Whitcomb et al. (2020). This tracer was calibrated to match free–free radio continuum emission using a similar data set based on SINGS extranuclear spectra. The SFR surface density $\Sigma_{\rm SFR}$ is given by

$$\left(\frac{\Sigma_{\rm SFR}}{M_\odot\ {\rm yr}^{-1}\,{\rm kpc}^{-2}}\right) = 6.2 \times 10^{-3.25} \times (\Sigma{\rm Ne})^{0.87} \times ([{\rm O/H}])^{-0.48}, \quad (2)$$

where constants are taken from their Table 5, $\Sigma$ Ne is the sum of the 15.6 $\mu$m [Ne III] and the 12.8 $\mu$m [Ne II] integrated line intensities in $10^{-7}\,{\rm W\,m^{-2}\,sr^{-1}}$, and [O/H] is from KK04-calibrated $12 + \log_{10}$[O/H] metallicities. We use this as our reference SFR tracer and refer to it as the $\Sigma$Ne and Z tracer. This tracer is based on fine structure lines from high ionization potential ions ($I > 13.6$ eV) and it is calibrated to free–free emission. Our reference tracer is then tracing ionizing photons generated by recent star-forming events ($\lesssim$5 Myr). Since our reference tracer responds to SF events on short timescales, an emission feature could be less correlated with our SFR if it responds directly to SF on longer timescales. The $\Sigma$Ne and Z tracer is also independent of dust emission and attenuation. It is crucial that our SF tracer is independent of any dust attenuation or obscuration correction, since our goal is to separate the correlation of various MIR observables with SF and gas separately.

We also include 21 regions that have detections of only one neon line, with the undetected line set to zero. Our results are not altered by including these regions and they typically have the lowest $\Sigma_{\rm SFR}$ in our sample but do not deviate from observed trends (see Figure 3). Equation (2) gives the SFR surface density $\Sigma_{\rm SFR}$, and we refer to this quantity as simply SF or SFR interchangeably for the remainder of this paper.

### 2.7. MIR Spectral Fitting

The extracted SL spectra were fit with the IDL program PAHFIT (Smith et al. 2007). We input heliocentric systemic velocities from the Lyon–Meudon Extragalactic Database (LEDA; Makarov et al. 2014) to correct for redshift variations. Constraints on the attenuation are weak with only SL coverage and previous studies of SINGS regions have found that, excepting specific cases such as edge-on targets, the attenuation in the MIR is generally minimal (Roussel et al. 2007; Smith et al. 2007), so we do not fit the attenuation. From PAHFIT we obtained the integrated intensities of emission features, as well as PAH, line, and continuum component spectra. The subfeatures of major PAH complexes are combined such that we obtain integrated intensities for the following bands and band complexes: 6.2, 7.7, 8.3, 8.6, 11.3, 12.0, 12.6, 13.6, and 14.2 $\mu$m. Uncertainties for these bands and emission lines are also returned by PAHFIT.

At present PAHFIT only operates on low-resolution spectra. To separate the underlying hot dust continuum from PAH emission in high-resolution SH spectra, we constructed a program similar to PAHFIT following Smith et al. (2007). We modeled fine structure and molecular line emission with Gaussian profiles. PAH features were modeled with the Drude emission profiles. We subtracted two linear estimates of the continuum between three points with no nearby emission features: one between 10 and 15 $\mu$m, and another between 15 and 18.2 $\mu$m. We then fit the spectrum from 10–15 $\mu$m as a linear combination of the individual Gaussian and Drude profiles using initial estimates and constraints for the parameters shown in Table 2 adopted from Smith et al. (2007), except for those that compose the 11.3 $\mu$m PAH complex. We find this complex is better fit by four Drude profiles in the higher spectral resolution SH data than the analogous two Drude profiles prescribed in Smith et al. (2007), which are designed for lower spectral resolution SL data. Notably, the 11.0 $\mu$m PAH feature can be fit distinctly from the rest of the complex at 11.3 $\mu$m, which we find is best described by PAH components at 11.2, 11.25, and 11.4 $\mu$m. The same technique is also applied to the spectral features between 15 and 19 $\mu$m.

We obtain uncertainties on the integrated intensities with Monte Carlo trials. In each trial, we perturb each point in the SH spectrum within its uncertainty assuming a Gaussian distribution, then run our fitting routine to find the integrated intensity of each emission feature. We then take the mean and standard deviation of 500 trial fits as the value and uncertainty of the feature. For convenience, we refer to this method and the program used to implement it as the Fitter for Aromatic, Atomic, and Molecular features in the Infrared (FARAMIR). Figure 1 (center) shows the FARAMIR fit for an SH observation of an extranuclear region in NGC 5194. In

**Table 2**
SH and LH Feature Initial Conditions

| Feature | Peak Wavelength | FWHM | Fractional FWHM |
|---|---|---|---|
| [S IV] | 10.511 | 0.100 | ⋯ |
| PAH 11.0 $\mu$m[a] | 11.01 | ⋯ | 0.012 |
| PAH 11.2 $\mu$m[a] | 11.21 | ⋯ | 0.008 |
| PAH 11.25 $\mu$m[a] | 11.25 | ⋯ | 0.012 |
| PAH 11.4 $\mu$m[a] | 11.40 | ⋯ | 0.032 |
| PAH 12.0 $\mu$m | 11.99 | ⋯ | 0.045 |
| $H_2$S(2) | 12.278 | 0.100 | ⋯ |
| Humphreys-$\alpha$ | 12.372 | 0.100 | ⋯ |
| PAH 12.6 $\mu$m | 12.62 | ⋯ | 0.042 |
| PAH 12.7 $\mu$m | 12.69 | ⋯ | 0.013 |
| [Ne II] | 12.813 | 0.100 | ⋯ |
| PAH 13.5 $\mu$m | 13.48 | ⋯ | 0.040 |
| PAH 14.2 $\mu$m | 14.19 | ⋯ | 0.025 |
| [Ne V] | 14.33 | 0.100 | ⋯ |
| [Ne III] | 15.555 | 0.140 | ⋯ |
| PAH 15.9 $\mu$m | 15.90 | ⋯ | 0.020 |
| PAH 16.5 $\mu$m | 16.45 | ⋯ | 0.014 |
| $H_2$S(1) | 17.035 | 0.140 | ⋯ |
| PAH 17.0 $\mu$m | 17.04 | ⋯ | 0.065 |
| PAH 17.4 $\mu$m | 17.375 | ⋯ | 0.012 |
| PAH 17.9 $\mu$m | 17.87 | ⋯ | 0.016 |
| [S III] | 18.713 | 0.140 | ⋯ |
| $H_2$S(0) | 28.221 | 0.020 | ⋯ |
| [S III] | 33.480 | 0.020 | ⋯ |
| [Si II] | 34.815 | 0.020 | ⋯ |

**Note.**
[a] Decomposition of PAH 11.3 $\mu$m complex at high spectral resolution differs from Smith et al. (2007).

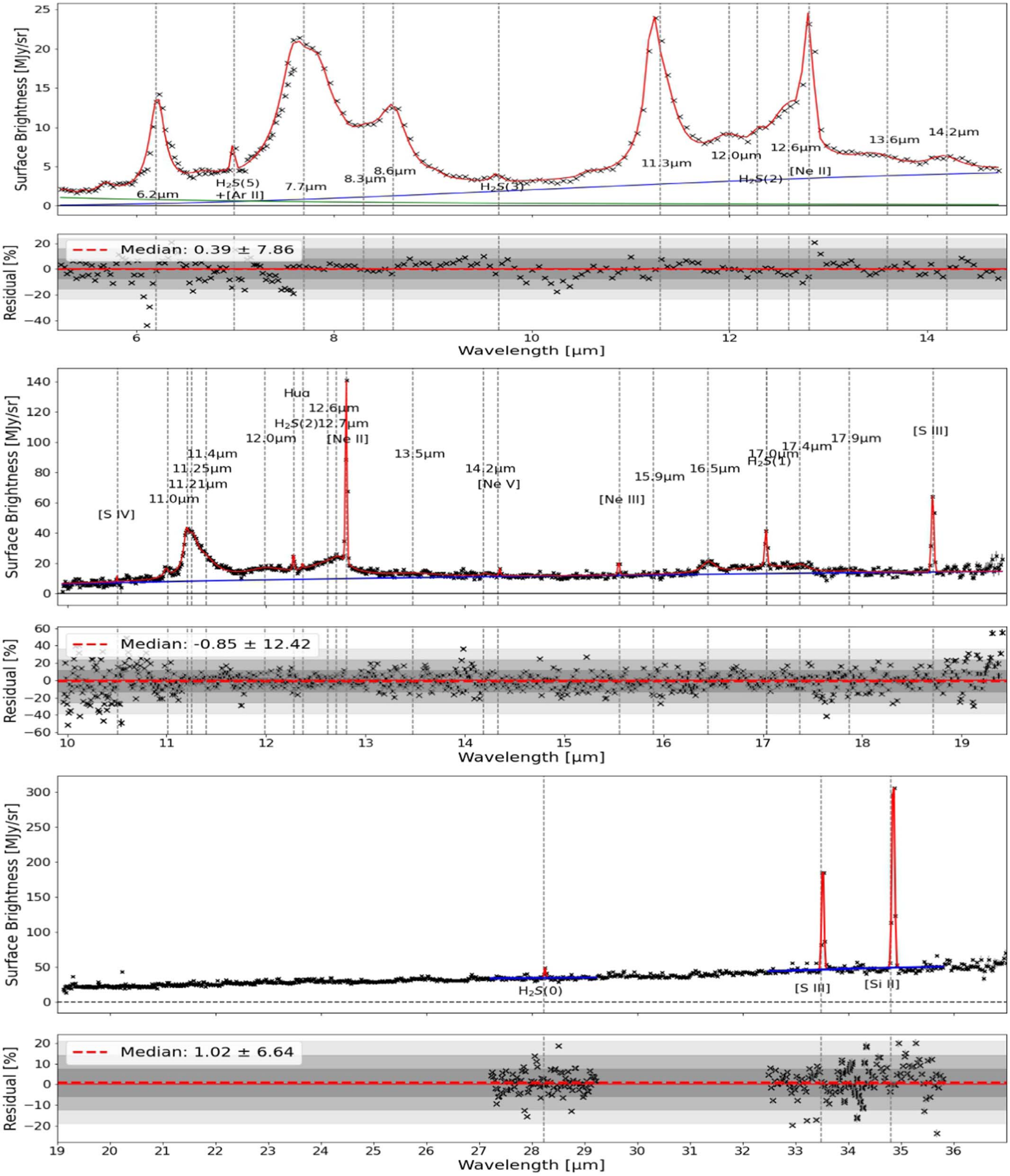

**Figure 1.** Example spectra of SINGS extranuclear region 0 in NGC 5194. (top) PAHFIT result for the SL spectrum, (center) FARAMIR result for the SH spectrum, and (bottom) Gaussian fitting result for the LH spectrum. The central wavelengths of the PAH features and emission lines we attempt to fit with Drude or Gaussian profiles are indicated by vertical dashed lines. The feature fit is shown in red, the stellar continuum fit is shown in green, and the dust continuum fit is shown in blue. The bottom panel of each spectrum shows the remaining residual percent difference when our fit is subtracted, where the dashed line indicates the median value and the gray bands denote once, twice, and thrice the standard deviation of these residuals. We note the continuum level typically differs between the three orders, but SL is properly calibrated so this difference does not affect our analysis.

Appendix A.2 we verify the accuracy of FARAMIR by comparison with PAHFIT results for analogous SL spectra. FARAMIR reproduces the relevant PAH feature and emission line intensities but overestimates by ∼20% or less on average relative to PAHFIT. This overestimation is likely a result of the simple linear continuum model used by FARAMIR, whereas PAHFIT uses a set of several blackbody functions to model the dust continuum.

The LH spectra contain only three emission lines that can be fit in most regions: $H_2$ S(0) at 28.2 $\mu$m, [S III] at 33.5 $\mu$m, and [Si II] at 34.8 $\mu$m. Beyond 35 $\mu$m the S/N is too low to reliably fit emission lines such as [Ne III] at 36.0 $\mu$m. The other notable lines in the LH spectral range, [O IV] at 25.91 $\mu$m and [Fe II] at 25.99 $\mu$m, are too faint relative to the LH periodic continuum noise (*scalloping*[5]) to be fit separately in most regions. We fit the LH spectra with three individual Gaussian profiles: for each line we crop the spectrum to 1 $\mu$m on either side of the peak wavelength, then fit a linear continuum using the average surface brightness between 1 and 0.5 $\mu$m on either side of the peak wavelength. This continuum is subtracted from the cropped spectrum and we follow the same Monte Carlo procedure outlined above to obtain integrated intensities and their uncertainties.

We quantify the detection threshold of our fits to emission features in the SH and LH spectra using the residual scatter after the fit is subtracted from the original data. We calculate the standard deviation of the residual difference and compare it with the fitted amplitudes determined by our method. We define our non-detection threshold as thrice the standard deviation of the residuals, such that any Drude or Gaussian with a fitted amplitude below this threshold is considered a non-detection.

We find very few detections for the 15.9 and 17.9 $\mu$m PAH features, likely because these are intrinsically weak and immediately near our assumed continuum points at 15 and 18.2 $\mu$m. Our linear continuum assumption for fitting SH spectral features is sufficient for most bright spectra from star-forming regions, but some have a significantly nonlinear continuum and a small but non-negligible portion of the PAH features can be removed by the linear fit. Likely as a result of this linear fit, we find less than fifteen $3\sigma$ detections of the 13.5 and 14.2 $\mu$m PAH features with our FARAMIR method from SH spectra, but this is not the case for these same PAH features as measured by PAHFIT from SL spectra. When necessary we use the PAHFIT results from SL spectra for these low S/N PAH features in our analysis. We also find less than fifteen detections for the 17.9 $\mu$m PAH feature. All major emission features in the SL spectral overlap with SH measured with PAHFIT and FARAMIR respectively are in good agreement (see the Appendix).

### 2.8. Synthetic Photometry

In order to study the contributions of PAH features, emission lines, and dust continuum to the IRAC4 8 $\mu$m and WISE3 12 $\mu$m photometric bands, we perform synthetic photometry on decomposed IRS spectra. The WISE3 12 $\mu$m photometric band is sensitive to emission from 7.2–16.4 $\mu$m. This range overlaps with the IRAC4 band at 8 $\mu$m, which collects from 6.2–9.7 $\mu$m. The SL1 and SL2 combined spectra span the full IRAC4 band, but the SL spectral range ends at 14.7 $\mu$m. For the remaining portion of the WISE3 wavelength coverage outside the IRS SL range, we append the 14.7–16.4 $\mu$m range of the corresponding SH spectrum. The continuum level differs between SL and SH, so we subtract a linear continuum from the SH as determined by the average brightness at 10.1 and 14.4 $\mu$m. We then match the continuum-subtracted SH spectrum to the SL spectrum based on the average SL brightness at these same wavelengths. The final stitched spectra are composed of the full SL spectrum from 5.2–14.7 $\mu$m and the continuum-adjusted SH spectrum from 14.7–16.4 $\mu$m. These combined spectra are only used for our synthetic WISE3 photometry analysis.

We use PAHFIT to decompose the SL1 and SL2 spectra into PAH, starlight continuum, dust continuum, and line emission contributions in the 120 non-AGN spectra. For the wavelengths longer than the SL coverage (greater than 14.7 $\mu$m) we consider the appended 14.7–16 $\mu$m portion of the SH spectrum to be continuum. The WISE3 bandpass is sensitive out to about 16 $\mu$m and other than continuum, the 15.6 $\mu$m [Ne III] line is the only emission feature included between 15 and 16 $\mu$m. We find it contributes a negligible amount to the total integrated flux from 15–16 $\mu$m compared to continuum emission for the majority of regions.

From the decomposed SL spectra we find the IRAC4 8 $\mu$m photometric band is dominated by emission from the PAH complex at 7.7 $\mu$m. But there is a roughly equal split between PAH and dust continuum portions for the full WISE3 12 $\mu$m band. Figure 2 shows the dust continuum contributes a larger fraction of emission to the WISE3 band than PAHs. The median fractional contribution to IRAC4 from PAHs is found to be 78%, that from continuum about 19%, and that from lines about 4%. The median fractional contribution from PAHs to WISE3 is 38%, that from continuum is 58%, and that from lines is about 4%. We note that the PAH contribution to WISE3 only exceeds 50% in a handful of regions.

### 2.9. Correlation Methods

We quantify the correlation between two emission features with the Spearman rank correlation coefficient. We also fit a power law to the data by ordinary least-squares using the logarithmic form:

$$\log_{10}(\mathrm{Y}) = m \times \log_{10}(\mathrm{X}) + \mathrm{b} \qquad (3)$$

where Y is CO [K km $\mathrm{s}^{-1}$] or $\Sigma_{\mathrm{SFR}}$ [$M_\odot$ $\mathrm{yr}^{-1}$ $\mathrm{kpc}^{-2}$], X is the comparison emission feature converted to surface brightness units $10^{-7}$ W $\mathrm{m}^{-2}$ $\mathrm{sr}^{-1}$ for all quantities (other than CO and $\Sigma_{\mathrm{SFR}}$), $m$ is the power-law slope, and $b$ is the power-law intercept.

We use a Monte Carlo method to account for uncertainties in each axis. In each of the trials we randomly offset each point according to its uncertainty assuming a Gaussian distribution. We simultaneously perform a bootstrap that randomly resamples the whole data set with replacement. Then we fit the constants of the power law and calculate the standard deviation and Spearman correlation coefficient. We complete 500 such trials and take the mean and standard deviation as the value of each quantity and its uncertainty.

Figure 3 shows the strongest correlations with CO and SF as an example. We also include examples of weaker correlations by showing the CO correlation for the feature best correlated with SF, and the SF correlation for the feature best correlated with CO. We note that the SF correlations have a different number of points than the CO correlations. The number of points varies based on the number of detections of each emission feature, but the maximum possible is 112 for SF and 67 for CO correlations. We find that restricting the set to include the same points in both correlations does not significantly alter our results.

[5] https://irsa.ipac.caltech.edu/data/SPITZER/SINGS/doc/sings_fifth_delivery_v2.pdf

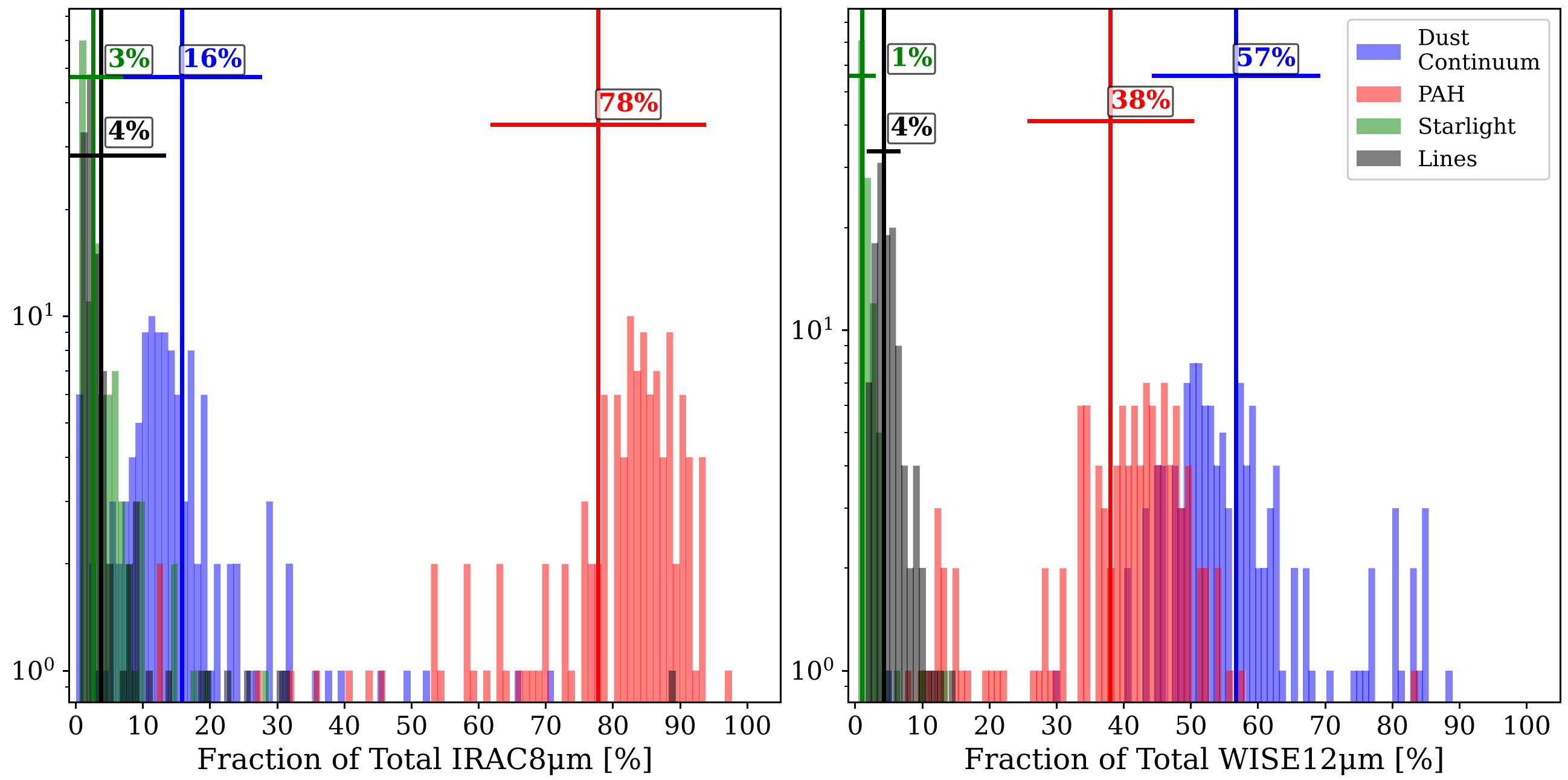

**Figure 2.** Fractional contribution of PAH (red), dust continuum (blue), starlight (green), and line emission (black) to total IRAC 8 $\mu$m (left) and WISE 12 $\mu$m photometry (right). The median and standard deviation of each distribution is indicated by the vertical and horizontal lines of the corresponding color.

The color gradient in Figure 3 shows the ratio of 15.6 $\mu$m [Ne III] to 12.8 $\mu$m [Ne II]. This ratio is a tracer of ionization parameter, which is itself strongly dependent on local properties such as the hardness of the radiation field, and therefore metallicity. The relation between [Ne III]/[Ne II] and metallicity has been investigated in many other studies (e.g Madden et al. 2006; Ho & Keto 2007; Xie & Ho 2019). Whitcomb et al. (2020) showed there is a strong, negative correlation between [Ne III]/[Ne II] and metallicity for apertures extracted from the same SINGS SL spectra used here.

## 3. Results

We measure the degree to which each feature traces CO or SFR by comparing their Spearman coefficients $\rho_{\rm CO}$ and $\rho_{\rm SFR}$, respectively. In our full sample, we find the Spearman coefficient of the correlation between SFR surface density $\Sigma_{\rm SFR}$ and CO emission (as a proxy for $\Sigma_{\rm H2}$) is about 0.4. Since our sample is focused exclusively on bright H II regions, this value represents the *minimum* $\Sigma_{\rm SFR}$ correlation. This $\Sigma_{\rm SFR}$–CO correlation is due in part to the K-S relation persisting on small scales. It is also due to our use of a short timescale reference SFR tracer, with which we cannot easily distinguish between a tracer of a varying SF history and a tracer of molecular gas from correlation alone.

Our goal is to find correlations that are significantly better, or worse, than the $\Sigma_{\rm SFR}$–CO relation. Table 3 shows the CO correlation statistics for each emission feature, and Table 4 shows them for the SFR correlations. The statistics from these tables (excluding the number of regions) are plotted in Figure 4.

Our results are visualized in Figure 4 where the correlation statistics with CO for each emission feature are shown on the horizontal axis of each plot and the corresponding statistics with $\Sigma_{\rm SFR}$ are on the vertical axes. Since SF and CO are themselves correlated in our sample of H II regions by the K-S relationship, we indicate the correlation coefficient between them in Figure 4 (top) with vertical and horizontal dashed lines at $\sim$0.4. The gray band shows the 1$\sigma$ range of the null hypothesis, determined by the standard deviation of correlation coefficients between two sets of random values. The null range varies based on the number of points (here 112 for the SFR axis and 67 for the CO axis), so correlation coefficients within this gray band are statistically equivalent to those from uncorrelated sets. The top-right plot shows the correlations which are stronger than those introduced solely by the K-S relationship from the top-left plot.

### *3.1. Breakdown in the K-S Relation Reveals SF-dominant and CO-dominant Tracers*

In our effort to quantify how well IR emission features trace SF and molecular gas, the breakdown of the K-S relationship on small scales provides some unique insights into the correlations. On the smallest scales, $H_2$ and ionized gas from SF are not colocated. However, they are often found nearby in the dense star-forming complexes studied here, so on large scales the correlation between SFR and CO is strong and it becomes difficult to distinguish between tracers of SF and tracers of CO (Schruba et al. 2010; Leroy et al. 2013; Schinnerer et al. 2019). Our regions sit at intermediate size scales ($\sim$100 s pc, but less than a kiloparsec), so there can be some distinction in the distributions. At the smallest sizes this distinction becomes even clearer.

Comparing the left and right plots in Figure 5, the dashed lines show the correlation of $\Sigma_{\rm SFR}$ with $\Sigma_{\rm H2}$ is significantly stronger for the 34 regions with an area greater than 0.4 kpc$^2$ compared to the correlation for the other 33 smaller regions ($\rho_s = 0.7$ versus 0.2). The dashed lines for the small regions

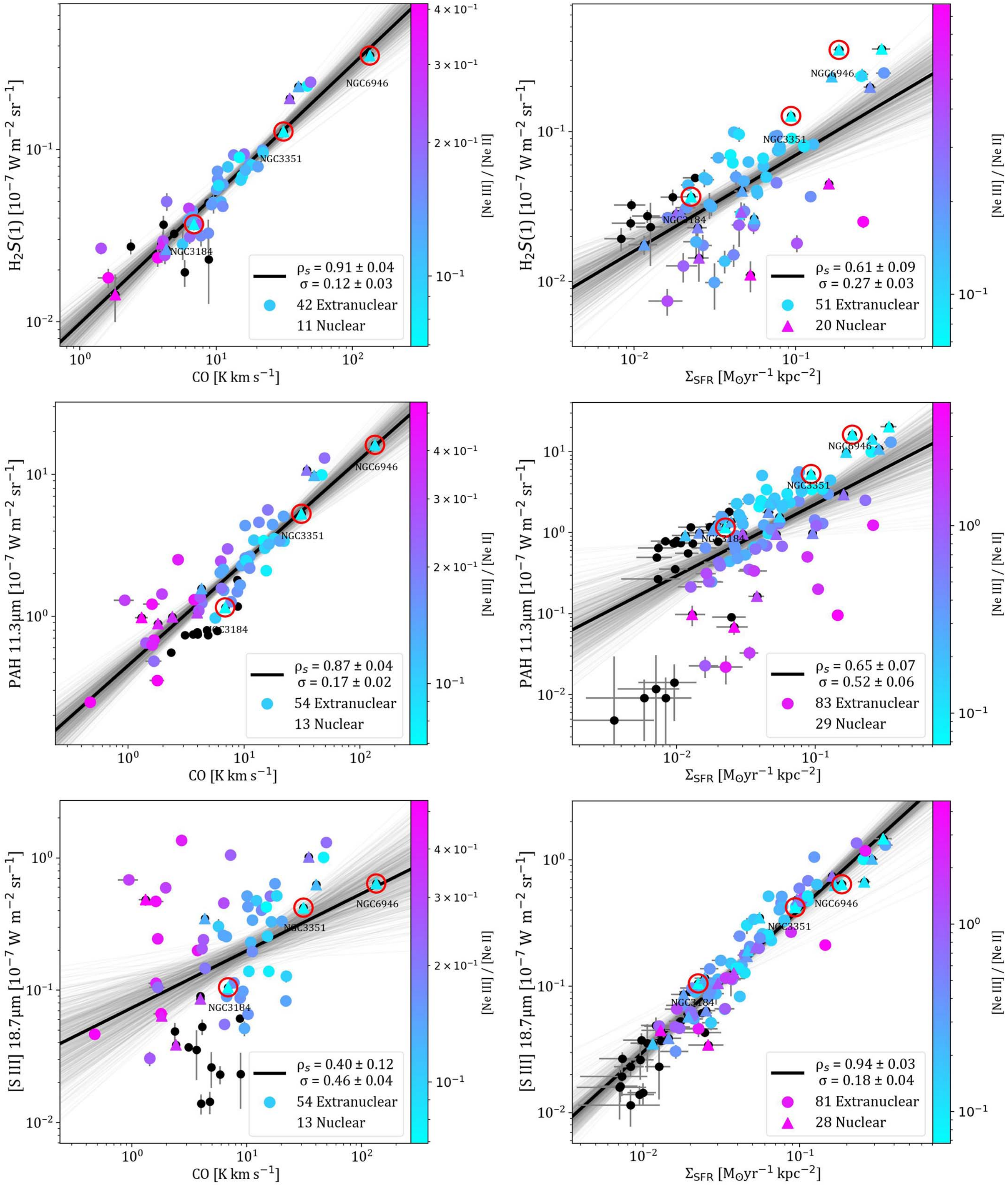

**Figure 3.** Examples of strong and matching weak correlations with CO and SF: the $H_2$ S(1) rotational line and CO $J = (2 - 1)$ emission (top left), $H_2$ S(1) and $\Sigma_{SFR}$ (top right), the 11.3 $\mu$m PAH complex and CO emission (center left), PAH 11.3 $\mu$m and $\Sigma_{SFR}$ (center right), the [S III] line at 18.7 $\mu$m and CO emission (bottom left), and [S III] 18.7 $\mu$m and $\Sigma_{SFR}$ (bottom right). The color gradient indicates the ratio of 15.6 $\mu$m [Ne III] to 12.8 $\mu$m [Ne II] and black points lack detection of one of these lines. The three red circles denote nuclear regions with $\alpha_{CO}$ significantly different than the Milky Way value (see Section 4.2).

**Table 3**
CO Correlation Statistics

| Feature | Correlation Coefficient $\rho_s$ | Standard Deviation $\sigma$ (dex) | Power-law Slope $m$ | Power-law Intercept $b$ | Number of Regions |
|---|---|---|---|---|---|
| WISE1 3.4 $\mu$m | $0.71 \pm 0.07$ | $0.25 \pm 0.02$ | $0.65 \pm 0.08$ | $0.50 \pm 0.08$ | 63 |
| IRAC1 3.6 $\mu$m | $0.72 \pm 0.07$ | $0.24 \pm 0.02$ | $0.67 \pm 0.08$ | $0.49 \pm 0.08$ | 63 |
| IRAC2 4.5 $\mu$m | $0.73 \pm 0.07$ | $0.24 \pm 0.02$ | $0.67 \pm 0.08$ | $0.25 \pm 0.08$ | 63 |
| WISE2 4.6 $\mu$m | $0.73 \pm 0.07$ | $0.23 \pm 0.02$ | $0.65 \pm 0.08$ | $0.21 \pm 0.08$ | 63 |
| IRAC3 5.8 $\mu$m | $0.85 \pm 0.05$ | $0.19 \pm 0.02$ | $0.76 \pm 0.07$ | $0.67 \pm 0.08$ | 63 |
| PAH 6.2 $\mu$m SL | $0.84 \pm 0.05$ | $0.20 \pm 0.02$ | $0.75 \pm 0.07$ | $-0.22 \pm 0.07$ | 67 |
| $H_2S(5)$+[Ar II] 6.9 $\mu$m | $0.65 \pm 0.09$ | $0.36 \pm 0.06$ | $0.70 \pm 0.12$ | $-1.67 \pm 0.13$ | 66 |
| PAH 7.7 $\mu$m SL | $0.87 \pm 0.04$ | $0.19 \pm 0.02$ | $0.79 \pm 0.06$ | $0.26 \pm 0.06$ | 67 |
| IRAC4 8 $\mu$m | $0.86 \pm 0.04$ | $0.19 \pm 0.02$ | $0.80 \pm 0.08$ | $0.87 \pm 0.08$ | 63 |
| IRAC4: cont. | $0.71 \pm 0.07$ | $0.22 \pm 0.02$ | $0.52 \pm 0.07$ | $2.48 \pm 0.07$ | 67 |
| IRAC4: PAH | $0.86 \pm 0.04$ | $0.19 \pm 0.02$ | $0.78 \pm 0.06$ | $3.13 \pm 0.06$ | 67 |
| PAH 8.3 $\mu$m SL | $0.87 \pm 0.04$ | $0.19 \pm 0.03$ | $0.81 \pm 0.08$ | $-0.80 \pm 0.09$ | 67 |
| PAH 8.6 $\mu$m SL | $0.86 \pm 0.04$ | $0.19 \pm 0.02$ | $0.75 \pm 0.06$ | $-0.43 \pm 0.06$ | 67 |
| $H_2S(3)$ SL | $0.81 \pm 0.09$ | $0.29 \pm 0.10$ | $0.65 \pm 0.15$ | $-2.08 \pm 0.17$ | 65 |
| [S IV] 10.5 $\mu$m | $0.15 \pm 0.14$ | $0.41 \pm 0.07$ | $-0.05 \pm 0.17$ | $-1.34 \pm 0.14$ | 29 |
| PAH 11.0 $\mu$m SH | $0.62 \pm 0.17$ | $0.15 \pm 0.02$ | $0.50 \pm 0.12$ | $-1.35 \pm 0.14$ | 24 |
| PAH 11.3 $\mu$m SH | $0.87 \pm 0.04$ | $0.23 \pm 0.04$ | $0.87 \pm 0.09$ | $-0.45 \pm 0.10$ | 66 |
| PAH 11.3 $\mu$m SL | $0.87 \pm 0.04$ | $0.17 \pm 0.02$ | $0.73 \pm 0.06$ | $-0.35 \pm 0.06$ | 67 |
| WISE3 12 $\mu$m | $0.82 \pm 0.06$ | $0.19 \pm 0.02$ | $0.72 \pm 0.07$ | $0.57 \pm 0.08$ | 63 |
| WISE3: cont. | $0.75 \pm 0.07$ | $0.25 \pm 0.03$ | $0.70 \pm 0.08$ | $3.11 \pm 0.08$ | 67 |
| WISE3: PAH | $0.87 \pm 0.04$ | $0.18 \pm 0.02$ | $0.78 \pm 0.06$ | $2.95 \pm 0.06$ | 67 |
| PAH 12.0 $\mu$m SH | $0.80 \pm 0.10$ | $0.12 \pm 0.01$ | $0.85 \pm 0.07$ | $-0.98 \pm 0.08$ | 30 |
| PAH 12.0 $\mu$m SL | $0.91 \pm 0.03$ | $0.16 \pm 0.02$ | $0.81 \pm 0.05$ | $-1.00 \pm 0.06$ | 67 |
| $H_2S(2)$ SH | $0.78 \pm 0.08$ | $0.21 \pm 0.06$ | $0.70 \pm 0.10$ | $-2.18 \pm 0.13$ | 47 |
| $H_2S(2)$ SL | $0.76 \pm 0.11$ | $0.29 \pm 0.09$ | $0.68 \pm 0.14$ | $-2.16 \pm 0.15$ | 67 |
| Hu$\alpha$ | $0.23 \pm 0.19$ | $0.25 \pm 0.08$ | $0.16 \pm 0.13$ | $-1.89 \pm 0.12$ | 29 |
| PAH 12.6 $\mu$m SH | $0.89 \pm 0.04$ | $0.25 \pm 0.11$ | $0.99 \pm 0.09$ | $-0.76 \pm 0.12$ | 47 |
| PAH 12.6 $\mu$m SL | $0.88 \pm 0.03$ | $0.17 \pm 0.02$ | $0.80 \pm 0.06$ | $-0.67 \pm 0.06$ | 67 |
| [Ne II] 12.8 $\mu$m SH | $0.66 \pm 0.08$ | $0.31 \pm 0.03$ | $0.65 \pm 0.09$ | $-0.99 \pm 0.10$ | 67 |
| [Ne II] 12.8 $\mu$m SL | $0.65 \pm 0.09$ | $0.32 \pm 0.03$ | $0.67 \pm 0.10$ | $-1.11 \pm 0.11$ | 67 |
| PAH 13.6 $\mu$m SL | $0.85 \pm 0.04$ | $0.23 \pm 0.03$ | $0.87 \pm 0.07$ | $-1.46 \pm 0.08$ | 66 |
| PAH 14.2 $\mu$m SL | $0.72 \pm 0.12$ | $0.37 \pm 0.08$ | $0.79 \pm 0.17$ | $-1.62 \pm 0.20$ | 67 |
| [Ne III] 15.6 $\mu$m | $0.11 \pm 0.09$ | $0.38 \pm 0.04$ | $0.00 \pm 0.12$ | $-1.04 \pm 0.12$ | 56 |
| $H_2S(1)$ | $0.91 \pm 0.04$ | $0.13 \pm 0.03$ | $0.75 \pm 0.06$ | $-2.01 \pm 0.07$ | 53 |
| $\Sigma$ PAH$_{17\mu m}$ | $0.85 \pm 0.07$ | $0.33 \pm 0.08$ | $1.19 \pm 0.21$ | $-1.63 \pm 0.28$ | 36 |
| [S III] 18.7 $\mu$m | $0.39 \pm 0.12$ | $0.46 \pm 0.04$ | $0.43 \pm 0.13$ | $-1.13 \pm 0.13$ | 67 |
| WISE4 22 $\mu$m | $0.68 \pm 0.09$ | $0.31 \pm 0.03$ | $0.75 \pm 0.11$ | $0.67 \pm 0.12$ | 63 |
| MIPS1 24 $\mu$m | $0.69 \pm 0.08$ | $0.32 \pm 0.03$ | $0.76 \pm 0.10$ | $0.60 \pm 0.11$ | 67 |
| $H_2S(0)$ | $0.85 \pm 0.05$ | $0.15 \pm 0.02$ | $0.64 \pm 0.05$ | $-2.33 \pm 0.05$ | 58 |
| [S III] 33.5 $\mu$m | $0.47 \pm 0.11$ | $0.41 \pm 0.03$ | $0.50 \pm 0.12$ | $-1.08 \pm 0.13$ | 67 |
| [Si II] 34.8 $\mu$m | $0.73 \pm 0.07$ | $0.25 \pm 0.02$ | $0.66 \pm 0.08$ | $-1.10 \pm 0.08$ | 67 |
| PACS 70 $\mu$m | $0.71 \pm 0.08$ | $0.28 \pm 0.03$ | $0.75 \pm 0.10$ | $1.25 \pm 0.10$ | 63 |
| PACS 100 $\mu$m | $0.85 \pm 0.05$ | $0.19 \pm 0.02$ | $0.92 \pm 0.08$ | $1.12 \pm 0.08$ | 46 |
| PACS 160 $\mu$m | $0.83 \pm 0.05$ | $0.17 \pm 0.02$ | $0.68 \pm 0.07$ | $1.14 \pm 0.07$ | 63 |
| SPIRE 250 $\mu$m | $0.87 \pm 0.04$ | $0.14 \pm 0.01$ | $0.61 \pm 0.06$ | $0.53 \pm 0.06$ | 63 |
| TIR | $0.69 \pm 0.08$ | $0.30 \pm 0.03$ | $0.77 \pm 0.10$ | $1.45 \pm 0.10$ | 67 |
| SFR | $0.45 \pm 0.12$ | $0.34 \pm 0.03$ | $0.40 \pm 0.11$ | $-1.68 \pm 0.11$ | 67 |
| Total neon | $0.55 \pm 0.10$ | $0.35 \pm 0.03$ | $0.54 \pm 0.10$ | $-0.81 \pm 0.11$ | 67 |
| Total [S III] | $0.44 \pm 0.12$ | $0.42 \pm 0.04$ | $0.46 \pm 0.12$ | $-0.80 \pm 0.13$ | 67 |
| [S III]+[S IV] | $0.34 \pm 0.12$ | $0.48 \pm 0.05$ | $0.37 \pm 0.13$ | $-1.04 \pm 0.14$ | 67 |
| Total $H_2S$ | $0.86 \pm 0.09$ | $0.27 \pm 0.08$ | $0.88 \pm 0.15$ | $-1.76 \pm 0.17$ | 67 |
| $H_2S(0+1+2)$ | $0.84 \pm 0.10$ | $0.29 \pm 0.09$ | $0.91 \pm 0.15$ | $-1.93 \pm 0.18$ | 63 |

**Note.** Ordered by wavelength, combinations of features at the bottom. Constants $\alpha$ and $\gamma$ apply to Equation (3). All correlations performed between the feature in $10^{-7}$ W m$^{-2}$ sr$^{-1}$ and CO in [K km s$^{-1}$], except $\Sigma_{\rm SFR}$ in [$M_\odot$ yr$^{-1}$ kpc$^{-2}$].

shown in Figure 5 (right) are less than the critical Spearman coefficient so the correlation between SF and CO in these regions is consistent with the null hypothesis. We note that the K-S correlation is likely weaker in general for our sample since our regions cover relatively small portions of many different galaxies.

While dividing the sample by size reveals the behavior of various tracers more clearly, it does introduce some biases in

**Table 4**
$\Sigma_{\rm SFR}$ Correlation Statistics

| Feature | Correlation Coefficient $\rho_s$ | Standard Deviation $\sigma$ (dex) | Power-law Slope $m$ | Power-law Intercept $b$ | Number of Regions |
|---|---|---|---|---|---|
| WISE1 3.4 $\mu$m | 0.48 ± 0.09 | 0.40 ± 0.03 | 0.65 ± 0.14 | 1.84 ± 0.20 | 97 |
| IRAC1 3.6 $\mu$m | 0.46 ± 0.09 | 0.45 ± 0.03 | 0.62 ± 0.14 | 1.74 ± 0.20 | 108 |
| IRAC2 4.5 $\mu$m | 0.51 ± 0.08 | 0.41 ± 0.03 | 0.63 ± 0.13 | 1.52 ± 0.19 | 108 |
| WISE2 4.6 $\mu$m | 0.54 ± 0.08 | 0.37 ± 0.03 | 0.70 ± 0.13 | 1.63 ± 0.18 | 97 |
| IRAC3 5.8 $\mu$m | 0.63 ± 0.07 | 0.44 ± 0.04 | 0.84 ± 0.15 | 2.25 ± 0.22 | 108 |
| PAH 6.2 $\mu$m SL | 0.68 ± 0.07 | 0.48 ± 0.07 | 0.87 ± 0.17 | 1.40 ± 0.24 | 109 |
| $H_2$S(5)+[Ar II] 6.9 $\mu$m | 0.62 ± 0.09 | 0.44 ± 0.07 | 0.75 ± 0.16 | −0.18 ± 0.22 | 106 |
| PAH 7.7 $\mu$m SL | 0.67 ± 0.08 | 0.50 ± 0.06 | 0.92 ± 0.18 | 1.95 ± 0.27 | 111 |
| IRAC4 8 $\mu$m | 0.64 ± 0.07 | 0.46 ± 0.04 | 0.90 ± 0.15 | 2.53 ± 0.22 | 107 |
| IRAC4: cont. | 0.69 ± 0.07 | 0.49 ± 0.13 | 0.83 ± 0.20 | 3.91 ± 0.27 | 112 |
| IRAC4: PAH | 0.67 ± 0.06 | 0.49 ± 0.05 | 0.95 ± 0.15 | 4.84 ± 0.23 | 112 |
| PAH 8.3 $\mu$m SL | 0.58 ± 0.08 | 0.47 ± 0.05 | 0.80 ± 0.13 | 0.77 ± 0.19 | 108 |
| PAH 8.6 $\mu$m SL | 0.66 ± 0.07 | 0.49 ± 0.08 | 0.86 ± 0.17 | 1.17 ± 0.25 | 110 |
| $H_2$S(3) SL | 0.51 ± 0.11 | 0.53 ± 0.09 | 0.54 ± 0.18 | −0.89 ± 0.23 | 105 |
| [S IV] 10.5 $\mu$m | 0.63 ± 0.10 | 0.37 ± 0.05 | 0.71 ± 0.18 | −0.40 ± 0.25 | 61 |
| PAH 11.0 $\mu$m SH | 0.60 ± 0.16 | 0.14 ± 0.02 | 0.50 ± 0.09 | −0.22 ± 0.11 | 25 |
| PAH 11.3 $\mu$m SH | 0.69 ± 0.06 | 0.44 ± 0.06 | 0.91 ± 0.11 | 1.38 ± 0.15 | 97 |
| PAH 11.3 $\mu$m SL | 0.65 ± 0.07 | 0.52 ± 0.07 | 0.89 ± 0.19 | 1.24 ± 0.27 | 112 |
| WISE3 12 $\mu$m | 0.76 ± 0.05 | 0.34 ± 0.04 | 0.95 ± 0.14 | 2.35 ± 0.19 | 97 |
| WISE3: cont. | 0.82 ± 0.05 | 0.31 ± 0.05 | 0.97 ± 0.13 | 4.91 ± 0.20 | 112 |
| WISE3: PAH | 0.67 ± 0.07 | 0.46 ± 0.05 | 0.91 ± 0.16 | 4.62 ± 0.23 | 112 |
| PAH 12.0 $\mu$m SH | 0.65 ± 0.12 | 0.21 ± 0.02 | 0.78 ± 0.11 | 0.89 ± 0.13 | 35 |
| PAH 12.0 $\mu$m SL | 0.60 ± 0.08 | 0.43 ± 0.06 | 0.76 ± 0.15 | 0.53 ± 0.23 | 109 |
| $H_2$S(2) SH | 0.64 ± 0.11 | 0.29 ± 0.06 | 0.60 ± 0.13 | −0.77 ± 0.17 | 64 |
| $H_2$S(2) SL | 0.62 ± 0.08 | 0.40 ± 0.07 | 0.67 ± 0.13 | −0.79 ± 0.18 | 110 |
| Hu$\alpha$ | 0.64 ± 0.13 | 0.24 ± 0.08 | 0.45 ± 0.12 | −1.20 ± 0.15 | 46 |
| PAH 12.6 $\mu$m SH | 0.60 ± 0.11 | 0.35 ± 0.06 | 0.85 ± 0.18 | 1.27 ± 0.23 | 56 |
| PAH 12.6 $\mu$m SL | 0.62 ± 0.07 | 0.50 ± 0.07 | 0.89 ± 0.17 | 1.00 ± 0.24 | 112 |
| [Ne II] 12.8 $\mu$m SH | 0.83 ± 0.05 | 0.31 ± 0.05 | 1.03 ± 0.13 | 0.81 ± 0.20 | 106 |
| [Ne II] 12.8 $\mu$m SL | 0.83 ± 0.05 | 0.34 ± 0.06 | 1.09 ± 0.13 | 0.77 ± 0.20 | 111 |
| PAH 13.5 $\mu$m SH | 0.76 ± 0.17 | 0.22 ± 0.06 | 1.33 ± 0.31 | 0.94 ± 0.27 | 11 |
| PAH 13.6 $\mu$m SL | 0.58 ± 0.08 | 0.47 ± 0.06 | 0.77 ± 0.15 | 0.18 ± 0.21 | 108 |
| PAH 14.2 $\mu$m SH | 0.77 ± 0.12 | 0.20 ± 0.05 | 0.86 ± 0.23 | −0.14 ± 0.22 | 13 |
| PAH 14.2 $\mu$m SL | 0.57 ± 0.10 | 0.49 ± 0.07 | 0.78 ± 0.17 | 0.03 ± 0.22 | 100 |
| [Ne V] 14.3 $\mu$m | 0.48 ± 0.23 | 0.26 ± 0.09 | 0.41 ± 0.22 | −1.41 ± 0.30 | 16 |
| [Ne III] 15.6 $\mu$m | 0.78 ± 0.05 | 0.27 ± 0.03 | 0.85 ± 0.10 | 0.09 ± 0.15 | 97 |
| $H_2$S(1) | 0.62 ± 0.09 | 0.27 ± 0.03 | 0.65 ± 0.10 | −0.51 ± 0.14 | 71 |
| $\Sigma$ PAH$_{17\ \mu m}$ | 0.58 ± 0.13 | 0.50 ± 0.08 | 1.10 ± 0.23 | 0.88 ± 0.28 | 49 |
| [S III] 18.7 $\mu$m | 0.94 ± 0.02 | 0.18 ± 0.04 | 1.13 ± 0.08 | 0.74 ± 0.11 | 109 |
| WISE4 22 $\mu$m | 0.85 ± 0.04 | 0.27 ± 0.04 | 1.12 ± 0.12 | 2.74 ± 0.18 | 97 |
| MIPS1 24 $\mu$m | 0.85 ± 0.05 | 0.29 ± 0.05 | 1.09 ± 0.14 | 2.62 ± 0.20 | 112 |
| $H_2$S(0) | 0.62 ± 0.09 | 0.23 ± 0.02 | 0.50 ± 0.08 | −1.09 ± 0.12 | 71 |
| [S III] 33.5 $\mu$m | 0.93 ± 0.04 | 0.19 ± 0.06 | 1.07 ± 0.11 | 0.73 ± 0.17 | 110 |
| [Si II] 34.8 $\mu$m | 0.83 ± 0.04 | 0.26 ± 0.02 | 0.92 ± 0.09 | 0.57 ± 0.13 | 109 |
| PACS 70 $\mu$m | 0.84 ± 0.05 | 0.28 ± 0.05 | 1.01 ± 0.14 | 3.17 ± 0.20 | 98 |
| PACS 100 $\mu$m | 0.80 ± 0.07 | 0.32 ± 0.06 | 1.00 ± 0.16 | 3.18 ± 0.24 | 74 |
| PACS 160 $\mu$m | 0.73 ± 0.06 | 0.35 ± 0.05 | 0.88 ± 0.15 | 2.76 ± 0.22 | 98 |
| SPIRE 250 $\mu$m | 0.65 ± 0.08 | 0.36 ± 0.05 | 0.77 ± 0.15 | 1.94 ± 0.21 | 98 |
| TIR | 0.77 ± 0.05 | 0.38 ± 0.05 | 1.02 ± 0.14 | 3.34 ± 0.21 | 112 |
| CO | 0.49 ± 0.12 | 0.45 ± 0.07 | 0.70 ± 0.19 | 1.72 ± 0.27 | 70 |
| Total neon | 0.94 ± 0.04 | 0.22 ± 0.09 | 1.10 ± 0.16 | 1.05 ± 0.23 | 112 |
| Total [S III] | 0.95 ± 0.02 | 0.15 ± 0.02 | 1.10 ± 0.06 | 1.05 ± 0.08 | 108 |
| [S III]+[S IV] | 0.93 ± 0.03 | 0.20 ± 0.05 | 1.11 ± 0.10 | 0.79 ± 0.14 | 109 |
| Total $H_2$S | 0.48 ± 0.11 | 0.56 ± 0.07 | 0.63 ± 0.18 | −0.31 ± 0.25 | 108 |
| $H_2$S(0+1+2) | 0.58 ± 0.10 | 0.43 ± 0.06 | 0.67 ± 0.15 | −0.29 ± 0.20 | 89 |

**Note.** Ordered by wavelength, combinations of features at the bottom. Constants $\alpha$ and $\gamma$ apply to Equation (3). All correlations performed between the feature in $10^{-7}$ W m$^{-2}$ sr$^{-1}$ and $\Sigma_{\rm SFR}$ in [$M_\odot$ yr$^{-1}$ kpc$^{-2}$], except CO in [K km s$^{-1}$]. Note the correlation coefficient for "Total neon" is not equal to 1 since our reference $\Sigma_{\rm SFR}$ includes a metallicity correction as well.

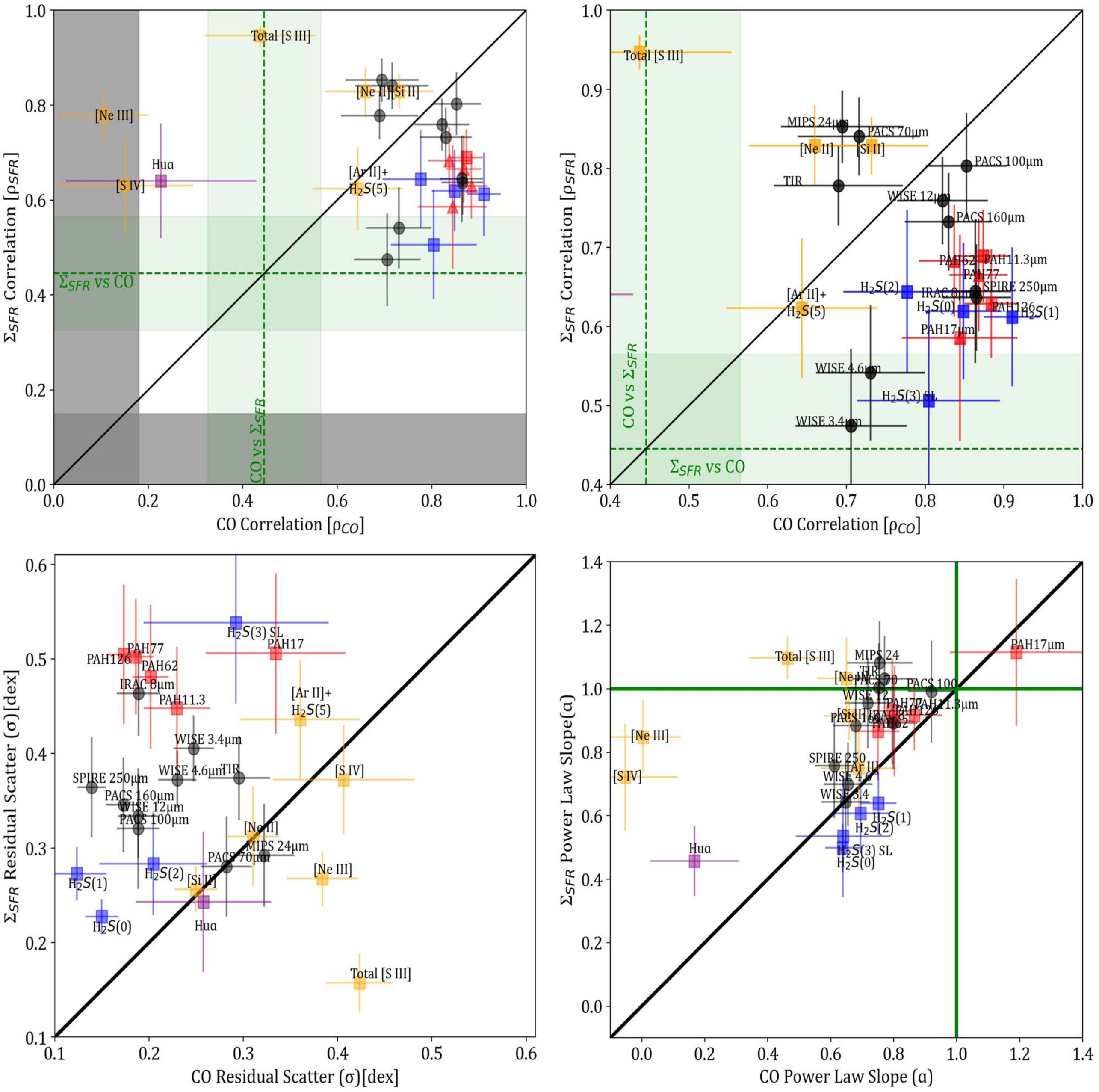

**Figure 4.** Summary statistics of IR emission features as $\Sigma_{SFR}$ and $\Sigma_{H2}$ tracers. The vertical and horizontal green lines in the top plots show the correlation coefficient $\rho_s$ and its $1\sigma$ uncertainty for CO as a function of $\Sigma_{SFR}$, and vice versa. The gray bands show the range where $\rho_s$ is statistically insignificant. PAH features are shown in red, fine structure lines in orange, $H_2$ rotational lines in blue, and photometric bands in black. The top-right panel shows the upper right quadrant of the top-left figure. (Bottom left) Same as top-left but for residual scatter in each fit instead of the Spearman coefficient, (bottom right) power-law slope of each fit with green lines indicating a slope of 1.

the selection as well. In particular, since the SINGS apertures are all of similar angular size, the smallest regions will be from the closest sources, which are more likely to be dwarf and/or lower metallicity galaxies. The relative positions of the various tracers in Figure 4 are mostly unaffected by looking at smaller size scales in Figure 5 (right), indicating that larger apertures simply have more $\Sigma_{SFR}$–CO overlap. We therefore draw our conclusions from the more statistically robust sample of all SF regions. We define an emission feature as SF dominant if it has greater $\rho_{SFR}$ than $\rho_{CO}$, and CO dominant for the opposite case.

### *3.2. Molecular Gas Correlations*

All major PAH features and $H_2$ rotational lines are strongly correlated with CO ($\rho_{CO} > 0.8$). From the other plots of Figure 4 we find the PAH features and $H_2$ rotational lines also have about the same sublinear power-law slope and the same amount of residual scatter in the CO correlation (both being within their $1\sigma$ uncertainties). The exception is the PAH 17 $\mu$m complex, which has a slope consistent with linear and a greater amount of residual scatter in the CO correlation compared to the shorter wavelength PAH features.

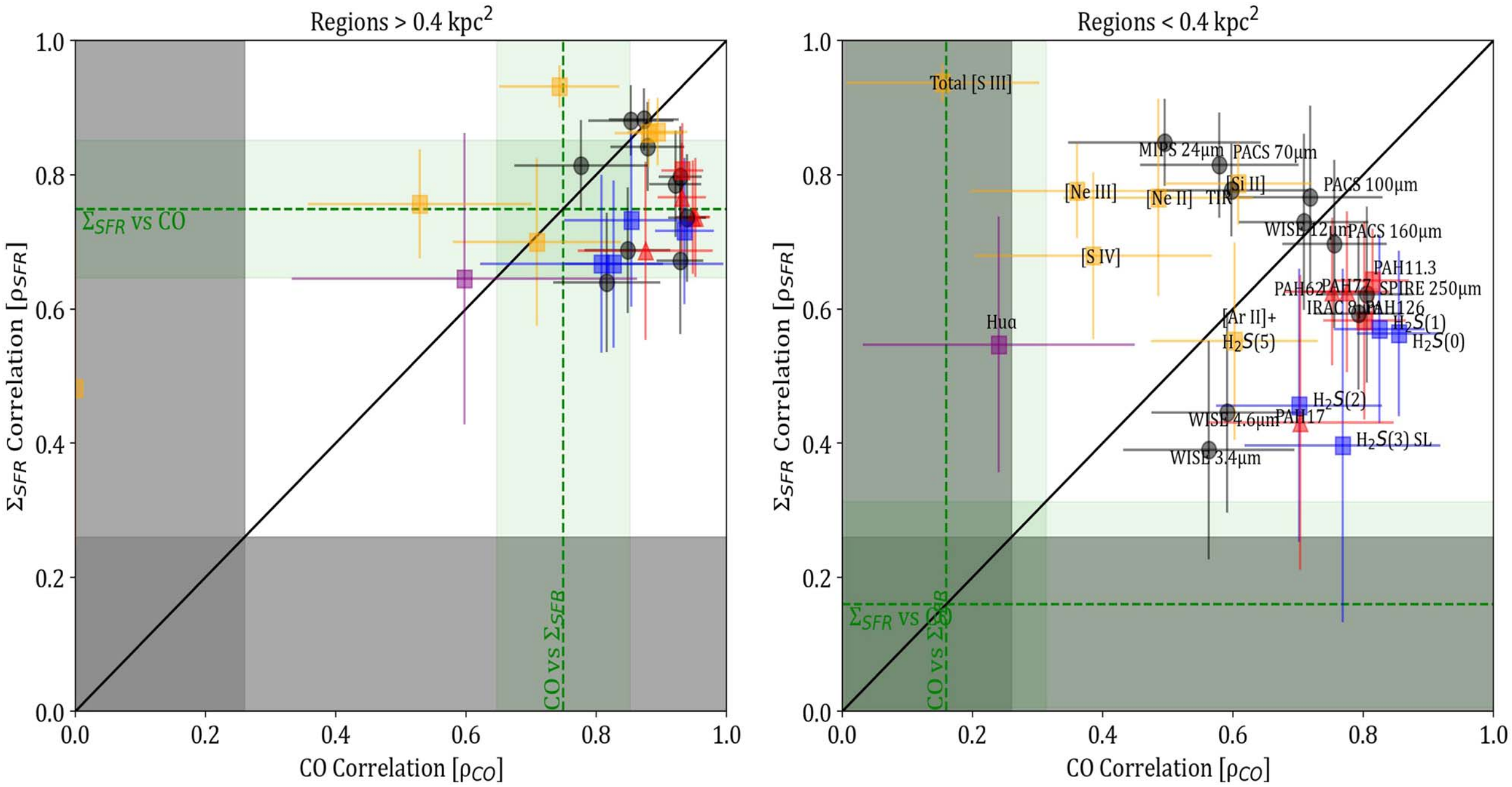

**Figure 5.** Same as Figure 4 (left) for 50% of regions based on physical aperture area (left $>0.4$ kpc$^2$; right $<0.4$ kpc$^2$).

All photometric bands from 3.4–250 $\mu$m are well correlated with CO. We find the bands at 24 and 70 $\mu$m have the lowest $\rho_{CO}$ (0.7) while the MIR 5.8 and 8 $\mu$m and FIR 100, 160, and 250 $\mu$m bands have the highest $\rho_{CO}$ (0.85). The TIR emission calculated from our photometric data has approximately the same $\rho_{CO}$ and $\rho_{SFR}$ as the 24 and 70 $\mu$m bands, implying that TIR for these regions reflects the warm dust in the vicinity of SF.

We find the weakest correlations with CO are fine structure lines from highly ionized species. For example, [Ne III] and [S IV] have $\rho_{CO}$ less than the critical Spearman coefficient 0.17, indicating the null hypothesis applies and they are not correlated with CO. These highly ionized lines and Humphreys-$\alpha$ are the only emission features with correlations weaker than the *minimum* K-S relation. This is possible because of our sample selection of regions focused on bright ionized gas associated with recent SF. This behavior is expected since emission from the most highly ionized gas is not correlated with molecular gas emission on small scales. We discuss this further in Section 4.3.

### 3.3. $\Sigma_{SFR}$ Correlations

All features and photometric bands have a stronger correlation with $\Sigma_{SFR}$ than the CO:$\Sigma_{SFR}$ correlation ($\rho_{SFR} > 0.4$, see Section 4). We find the [S III] lines and their sum, total [S III], are best correlated with $\Sigma_{SFR}$ with Spearman coefficient $\rho_{SFR}$ ∼0.95. Close behind at $\rho_{SFR} = 0.8$ are the 15.6 $\mu$m [Ne III] and 12.8 $\mu$m [Ne II] lines that compose our reference $\Sigma_{SFR}$, as well as the MIPS 24 $\mu$m band, which is a widely used SF indicator (Calzetti et al. 2007). The [Si II] line and PACS 70 $\mu$m emission also have $\rho_{SFR}$ ≈0.8. The weakest $\Sigma_{SFR}$ correlations are the $H_2$ S(3) line, and the short-wavelength photometric bands WISE1 3.4 $\mu$m and WISE2 4.6 $\mu$m with $\rho_{SFR} \sim 0.5$ for each. These features are less than 1$\sigma$ above the CO:SF correlation, and less than 1$\sigma$ below the PAH:SF correlation.

From Figure 4 (bottom left) we find the PAH:SF and $H_2$ S(3):SF correlations have the highest residual scatter. More than 1$\sigma$ below these are the IR continuum bands: WISE 3.4–12 $\mu$m, PACS 100, 160 $\mu$m, and SPIRE 250 $\mu$m. About 1$\sigma$ below these are the remaining $H_2$ rotational lines, fine structure lines, and the 24 and 70 $\mu$m IR bands. The [S III] lines or their sum have the lowest scatter of any SF correlation.

From Figure 4 (bottom right) we find most IR features have a slope with SFR that is consistent with linear to within 1$\sigma$. The exceptions are all sublinear ($m < 1$), the lowest being $m \sim 0.5$ for the $H_2$ rotational lines and Humphreys-$\alpha$, but we note the latter is a weakly detected line with few detections as a function of $\Sigma_{SFR}$.

### 3.4. Relative $\Sigma_{SFR}$ to $\Sigma_{H2}$ Tracing Potential

To quantify the difference between the SF- and CO-tracing features, we plot the difference between the correlation coefficients $\rho_{SFR} - \rho_{CO}$ from Figure 4 as a function of the observed wavelength in Figure 6.

Figure 6 (left) shows the difference in correlations $\rho_{SFR} - \rho_{CO}$ of each photometric band as a function of wavelength from WISE1 3.4 $\mu$m to SPIRE 250 $\mu$m. We find the emission from ∼20−70 $\mu$m (i.e., WISE 22 $\mu$m, MIPS 24 $\mu$m, and PACS 70 $\mu$m) is SF dominant. The WISE3 12 $\mu$m and PACS 100 $\mu$m bands are found to be slightly CO dominant, but not by greater than the 1$\sigma$ statistical uncertainty. All other bands, from WISE1 3.4 $\mu$m to IRAC4 8 $\mu$m, PACS 160 $\mu$m, and SPIRE 250 $\mu$m are CO dominant by greater than 1$\sigma$. The overall trend in this figure suggests a potential peak at about 40 $\mu$m, but we only have evidence that the continuum between 20 and 70 $\mu$m is likely a comparably strong SF tracer as the 24 and 70 $\mu$m emission.

We also use our synthetic photometry decomposition (see Section 2.8) to show the difference between SF and CO correlations for the continuum and PAH portions of the IRAC4 8 $\mu$m and WISE3 12 $\mu$m bands in Figure 6 (left). Notably, the continuum portion of the WISE band is SF dominant while the PAH portion is CO dominant, in good agreement with our expectations based on the behavior of the nearby 24 $\mu$m continuum and that of each individual PAH band in Figure 6. The distinction is less pronounced in the decomposed IRAC4 band. The continuum portion of both the IRAC and WISE bands better fits the general trend over wavelength seen in Figure 6 (left) than the total band with PAH emission included. This shows a clear difference in behavior between the PAH vibrational emission and the small, hot dust grain continuum in the MIR, which we discuss further in Section 4.

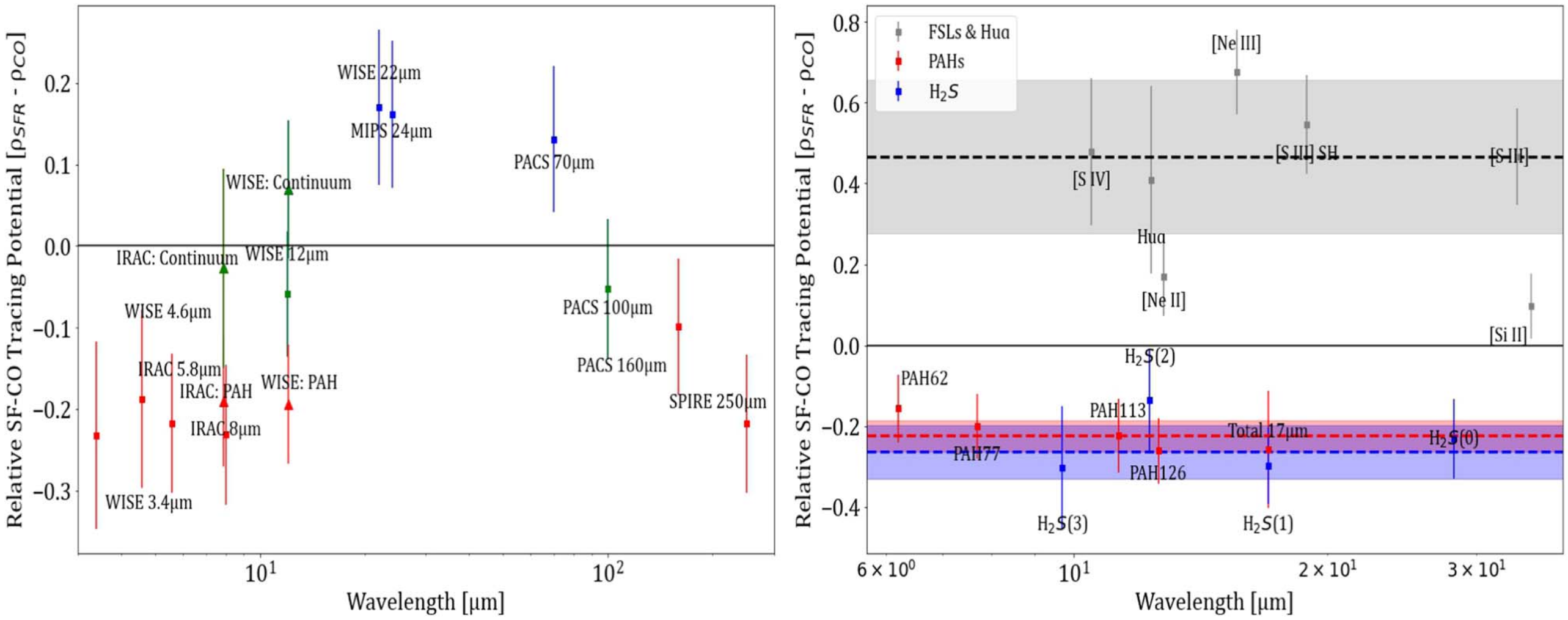

**Figure 6.** Difference in Spearman correlation coefficients ($\rho_s$) with SFR and CO as a function of wavelength for IR photometry (left) and MIR spectral features (right). The left plot shows CO-dominant continuum in red, SF-dominant continuum in blue, and neither CO dominant nor SF dominant by at least $1\sigma$ in green. The right plot shows PAH features in red, $H_2$ rotational lines in blue, and fine structure lines and Humphreys-$\alpha$ in gray. The horizontal dashed lines and bands show the median and standard deviation of each color.

From Figure 6 (right) we find the emission features from MIR spectra that are most CO dominant are the molecular gas rotational lines $H_2$ S(0) through $H_2$ S(3). The correlation coefficients of PAH features are remarkably similar to the $H_2$ rotational lines and to each other; they are less CO dominant by only about $1\sigma$ due to their stronger correlation with SF compared to the $H_2$ rotational lines. However, the residual scatter and power-law slopes vary significantly between PAH and $H_2$ rotational line correlations (see Figure 4, bottom; discussed further in Section 4.1.1).

We find the fine structure lines from ions with ionization potential greater than hydrogen—[Ne II], [Ne III], [S III], and [S IV]—are all SF dominant by at least $1\sigma$, with exception of the blend of the $H_2$ S(5) 6.91 $\mu$m and [Ar II] 6.99 $\mu$m lines, which traces SF and CO equally. Since we found the other $H_2$ lines are highly CO dominant, this blended line likely differs from the others since it includes a mix of CO-dominant $H_2$ emission in addition to the SF-dominant fine structure line emission from $Ar^+$. We also see the Humphreys-$\alpha$ line from ionized hydrogen is highly SF dominant, but this line is weakly detected and has large uncertainties. The remaining fine structure line from $Si^+$ has a lower ionization potential than hydrogen, and therefore may originate both in the ionized and neutral gas, but we find it is SF dominant by $\sim 1\sigma$.

We find both of the lines from $S^{2+}$ are highly SF dominant and behave indistinguishably from the reference SF tracer itself in Figures 4 and 5. This suggests either of the [S III] lines can be used as a single-parameter MIR tracer of star formation (see Section 4.3.1). The correlation is equally strong for each [S III] line individually as well as all combinations of summing them with the [S IV] line, as shown in Figure 7. For all individual and summed lines among [S IV], [S III] 18.7 $\mu$m, and [S III] 33.5 $\mu$m, the slope of their correlation with SF is about $1\sigma$ greater than linear ($m \sim 1.1$, see Table 4). There is a small trend in the scatter of these correlations with the ratio of 15.6 $\mu$m [Ne III] to 12.8 $\mu$m [Ne II]. However, including this ratio, or its physical correlate, metallicity (Madden et al. 2006; Whitcomb et al. 2020), as an additional feature in the regression results in a negligible improvement to the already tight correlation with $\Sigma_{\rm SFR}$, slightly reducing the scatter $\sigma$.

This behavior is distinct from the similar pair of neon MIR fine structure lines, 12.8 $\mu$m [Ne II] and 15.6 $\mu$m [Ne III]. The neon lines individually as a function of $\Sigma_{\rm SFR}$ have a strong metallicity dependence in the scatter, which is partially corrected by summing the two lines (see Zhuang et al. 2019; Whitcomb et al. 2020).

Using the fit constants from Table 4 in Equation (3), the single-line tracer of $\Sigma_{\rm SFR}$ based on the 18.7 $\mu$m [S III] line is given by

$$\left(\frac{\Sigma_{\rm SFR}}{M_\odot\ {\rm yr}^{-1}\,{\rm kpc}^{-2}}\right) = 10^{0.74\,\pm\,0.11} \times \left(\frac{[{\rm S\,III}]\,18.7\ \mu{\rm m}}{10^{-7}\ {\rm W\ m^{-2}\ sr^{-1}}}\right)^{1.13\,\pm\,0.08}. \quad (4)$$

## 4. Discussion

We find emission lines from ionized gas that are independent of dust content all have very strong correlations with our ionizing-photon-based $\Sigma_{\rm SFR}$ tracer. Dust-dependent emission features, given their strong correlation with CO, are mainly tracing the amount of dust but each has a varying degree of sensitivity to the local heating conditions. The correlations found in this study may arise from common responses to physical environment, e.g., radiative heating spectrum and intensity, or through spatial coupling of distinct environments, i.e., H II regions and molecular clouds. We found CO $J = (2-1)$ emission has the weakest correlation with $\Sigma_{\rm SFR}$. This is expected because our sample is focused on the brightest areas of star formation, so at the smallest scales molecular and ionized gas are spatially distinct and respond differently to local radiation conditions. The IR emission in and around star-forming regions is excited by UV photons, so the correlation of each IR feature with $\Sigma_{SFR}$ is stronger than the underlying $\Sigma_{SFR}$–CO relation because of this UV intensity correlation.

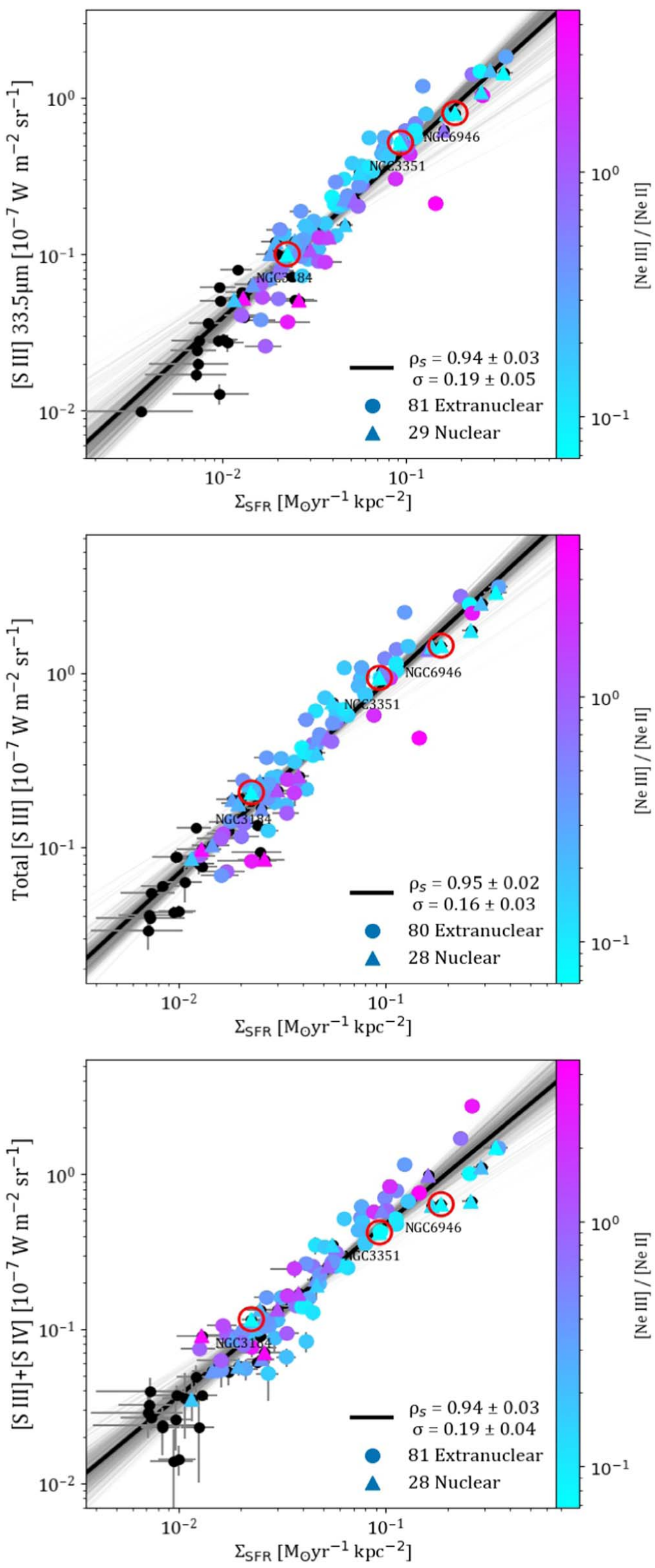

**Figure 7.** Integrated intensity of [S III] 33.5 $\mu$m (top), the sum of [S III] 18.7 $\mu$m and [S III] 33.5 $\mu$m (middle), and the sum of [S IV] 10.5 $\mu$m and [S III] 18.7 $\mu$m (bottom) as a function of $\Sigma_{SFR}$.

### 4.1. Mutual Correlation PAH:CO:$H_2$ Emission

The most CO-dominant tracers of Figure 6 (right) are found to be the rotational lines of $H_2$ followed closely by the PAH features. We find a strong mutual correlation between these three types of emission (Spearman coefficient >0.8); each $H_2$ line intensity is well correlated with both PAH and CO emission intensities, and the brightest PAH features are just as well correlated with both $H_2$ and CO emission. The correlation between PAH and $H_2$ emission has been well established in the literature for regions with negligible nonradiative heating (i.e., heating not driven by UV photons) such as the star-forming regions in this study (Roussel et al. 2007; Naslim et al. 2015; Smercina et al. 2018). In regions with significant nonradiative heating (e.g., shocks, turbulent dissipation), studies have found enhanced $H_2$ rotational relative to PAH emission (Ingalls et al. 2011). Spatial and intensity correlations between PAH and CO emission have also been observed in many previous and recent studies (Regan et al. 2006; Cortzen et al. 2019; Chown et al. 2021; Leroy et al. 2023).

In Section 4.1.1 we discuss the PAH and $H_2$ correlations with $\Sigma_{SFR}$ and CO in detail. The PAH and $H_2$ correlations with CO have about the same amount of residual scatter (except PAH 17 $\mu$m versus CO which has higher scatter). The PAH features, however, have more scatter than $H_2$ rotational lines do versus SF (except $H_2$ S(3)). The scatter in the PAH:SFR relation is dependent on metallicity, but metallicity does not introduce scatter in the PAH:CO relation (see center panels of Figure 3). All the $H_2$ lines have a sublinear slope as a function of both CO and SF, while the PAHs are approximately linear as a function of SF and the slope as a function of CO varies from sublinear for shorter wavelength features to superlinear for longer wavelength features.

The strong correlations between PAHs, CO, and $H_2$ rotational lines agree with expectations for emission generated by dense PDRs. In typical star-forming regions PAH vibrational bands are excited by UV photons, CO rotational lines are collisionally excited, and the $H_2$ rotational lines are excited by both collisions and UV pumping (Tielens 2008; Pereira-Santaella et al. 2014). The collisional excitation rates are set by the density and the gas temperature, which is the result of heating from the photoelectric effect. Photoelectric (PE) heating from PAHs and dust is the dominant heating mechanism in the warm portion of dense PDRs and it continues to dominate at optical depths where CO emits most strongly (Wolfire et al. 1993, 2022). The PE efficiency of PAHs is found to depend on the incident UV field (Croxall et al. 2012) and through the resulting photoelectrons the gas is heated indirectly by recent SF. Thus, the PAH, CO, and $H_2$ rotational line emission are all tracing the conversion of UV photons into excitation and collisions, which may explain the strong correlations in dense PDRs both spatially and in intensity.

We found PAHs and $H_2$ lines have a significantly weaker correlation with SF compared to the fine structure lines from ions. Comparing the correlations with SF for CO, $H_2$, and PAHs, we see $H_2$ is slightly less correlated with SF than PAHs. This agrees with a picture where the $H_2$ excitation is mainly collisional whereas PAHs respond directly to the local UV field. The warm portion of a dense PDR ($10^{2\text{-}3}$K) is ideal for excitation of the $H_2$ rotational lines. Closer to the H II regions the UV flux is sufficient to dissociate $H_2$(Roussel et al. 2007; Togi & Smith 2016). Togi & Smith (2016) conclude that the warmer portion of gas traced by the MIR $H_2$ rotational lines is typically about 15% of the total molecular gas mass. The strong correlations found in our results at small and large scales between each of the $H_2$ rotational integrated line intensities and CO $J = (2 - 1)$ surface brightness are evidence of both a spatial connection and a connection in how the warm and cold molecular gas emission lines are excited. The envelope of warm gas surrounding cold molecular clouds is bright in $H_2$ emission, but some of it is *CO dark* (Wolfire et al. 2010; Leroy et al. 2011; Smith et al. 2014; Lee et al. 2015; Madden et al. 2020). In addition, we found the $H_2$ and CO emission have a similar correlation strength with SF (with $H_2$ stronger than CO by about $1\sigma$). This may be evidence that the $H_2$ rotational emission is tracing a fairly constant fraction of the total $H_2$, consistent with the findings of Roussel et al. (2007) and Togi & Smith (2016) for the SINGS sample.

The PAH emission features require UV excitation, and we find they have a better correlation with $\Sigma_{SFR}$ compared to $H_2$ rotational lines. PAHs are relatively large molecules and/or small dust grains, and it is theorized that high-energy photons may destroy them (Micelotta et al. 2010b; Chastenet et al. 2019). Therefore, we do not expect PAH emission to be spatially correlated with ionized gas on the smallest scales of individual H II regions ($\lesssim$10 pc). The PAH features are CO dominant, but the correlation PAH:SF is slightly stronger than $H_2$:SF (though by less than the $1\sigma$uncertainty) while PAH:CO is the same as $H_2$:CO. Unlike the low-$J$ $H_2$ rotational lines, PAH vibrational excitation is more dependent on radiation hardness (Draine et al. 2021), which may account for the small difference in SF correlation for PAH and $H_2$ emission.

Since PAH and $H_2$ rotational emission are both spatially correlated with molecular gas but only PAH emission is directly dependent on the UV intensity, we might expect the PAH:SF correlation to be significantly stronger than $H_2$:SF, but this is not what we observe. The large scatter in the PAH:SF correlation has a strong dependence on the ratio [Ne III]/[Ne II], but the scatter in $H_2$:SF has a negligible dependence on this ratio. This suggests PAH emission is more sensitive to changes in metallicity than $H_2$ emission, which has been established in many previous studies (Engelbracht et al. 2005; Madden et al. 2006; Gordon et al. 2008; Muñoz-Mateos et al. 2009; Sandstrom et al. 2012; Lai et al. 2020; Zang et al. 2022). The ratio [Ne III]/[Ne II] tracks changes in radiation hardness and metallicity, so the scatter in the PAH:SF correlation is likely driven by variations in these properties. PAHs respond directly to UV photons, but the resulting emission bands have a dependence on the abundance of PAHs and their size and charge distributions, which in turn are dependent on local environmental properties such as metallicity or radiation hardness (Draine et al. 2021). Our sample is an ensemble of star-forming regions from many different galaxies, so these dependencies on local conditions are likely contributing to the scatter that weakens the correlation between PAH emission and $\Sigma_{SFR}$.

#### 4.1.1. Connections between Individual PAH and $H_2$ Correlations with CO and $\Sigma_{SFR}$

In Figure 8 (top left) we show our power-law fit with CO for all four studied $H_2$ emission lines. We find the power-law slope of each correlation is approximately the same (about 0.7) but the intercept varies between the lines, in order from dimmest: $H_2$ S(0), (2), (3), and S(1) is the brightest of the four. The trend between these warm molecular gas tracers and CO is found to be nonlinear and the brightest of the warm gas lines S(1) has the strongest correlation with CO, and with all the major PAH features as well. Figure 8 (top right) shows these same correlations with $\Sigma_{SFR}$ for comparison.

The lack of variation in the slope of each correlation in Figure 8 (top left) suggests that the heating contribution by UV pumping does not change between the lines. If UV pumping was significant, we would expect the slope as a function of $\Sigma_{SFR}$ for higher-energy rotational lines, such as $H_2$ S(3), to be greater than the slope of the ground state S(0) line. Instead, we find the slopes are approximately equal for all four rotational lines ($\sim$0.55 $\pm$ 0.1). Since the correlation with CO for each line is also much stronger, we conclude direct excitation of $H_2$ via UV pumping is unlikely to be dominant compared to collisional excitation for the $H_2$ rotational lines in our sample. This agrees with the analysis of the SINGS sample by Roussel et al. (2007).

We also include the correlations for the four brightest PAH features in Figure 8 (bottom). In contrast with the $H_2$:CO correlations, the power-law slope increases for CO correlations with longer wavelength PAH features, ranging from 0.8 at 7.7 $\mu$m, 0.9 at 11.3 $\mu$m, and 1.2 at 17 $\mu$m (see Table 4). In general, the PAH features at longer wavelengths originate more from larger PAHs, while shorter wavelength PAH features are more effectively emitted by small PAHs (Draine & Li 2007; Maragkoudakis et al. 2020; Draine et al. 2021). The longer wavelength features are brighter relative to the 7.7 $\mu$m feature at higher CO brightness, which may suggest PAH growth and/or more neutral PAHs in molecular gas regions. PAH growth in molecular clouds has been proposed by previous studies in the literature (Sandstrom et al. 2010; Chastenet et al. 2019). The increase in the 17/7.7 $\mu$m ratio could also result from a decrease in the average hardness of the radiation field as a function of CO. PAHs exposed to harder radiation fields, i.e., with more UV relative to lower-energy photons, emit significantly more in shorter relative to longer wavelength features (Draine et al. 2021). If dust content is higher in regions with more CO, more of the photons will be attenuated which could decrease the average hardness of the radiation field that is exciting the PAHs, potentially accounting for the enhanced 17/7.7 $\mu$m PAH ratio as a function of CO emission.

In Figure 8 (bottom right) we find the slope of the correlation with $\Sigma_{SFR}$ is the same within $1\sigma$ for each major PAH feature. This implies that band strength variations are not driven by changes in the incident UV field intensity. The fact that the slope as a function of $\Sigma_{SFR}$ does not vary between PAH features could be explained if the radiation field hardness is not correlated with $\Sigma_{SFR}$, or if the PAH population is changing in such a way that negates the expected band variations. If it is the PAH population changing, then the size and charge distribution must be shifting to larger and more neutral PAHs as a function of $\Sigma_{SFR}$. This is possible if small PAHs are destroyed or coagulated into larger PAHs, as the PAH:CO slopes could be interpreted to suggest.

### 4.2. CO-to-$H_2$ Conversion Factor

Throughout this work, we have studied correlations with CO $J = (2 - 1)$ emission. The issues with converting to molecular gas mass with a CO-to-$H_2$ conversion factor are explored in-depth in Bolatto et al. (2013). The CO-to-$H_2$ conversion factor is generally used with CO $J = (1 - 0)$ emission, but our results are based on correlations with CO $J = (2-1)$ emission. CO $J = (2-1)$ can be converted to CO $J = (1-0)$ with a ratio $R_{21}$ that is ∼0.65 on average and can vary between ∼0.4 and 1.0 (Leroy et al. 2022). This introduces some uncertainty in translating our correlations with CO $J = (2-1)$ emission into correlations with $M_{H2}$. However, $R_{21}$ is correlated with $\Sigma_{SFR}$, so we expect the values for our sample to be relatively constant and near the high end of this range ($R_{21} \sim 1.0$). The CO-to-$H_2$ conversion factor also has a strong dependence on metallicity (Wolfire et al. 2010; Leroy et al. 2011; Glover & Clark 2012) and on CO excitation and optical depth (Israel 2020; Teng et al. 2022). Our sample of H II regions in nearby galaxies has a relatively small range of metallicities, with the exception of a few dwarf galaxies.

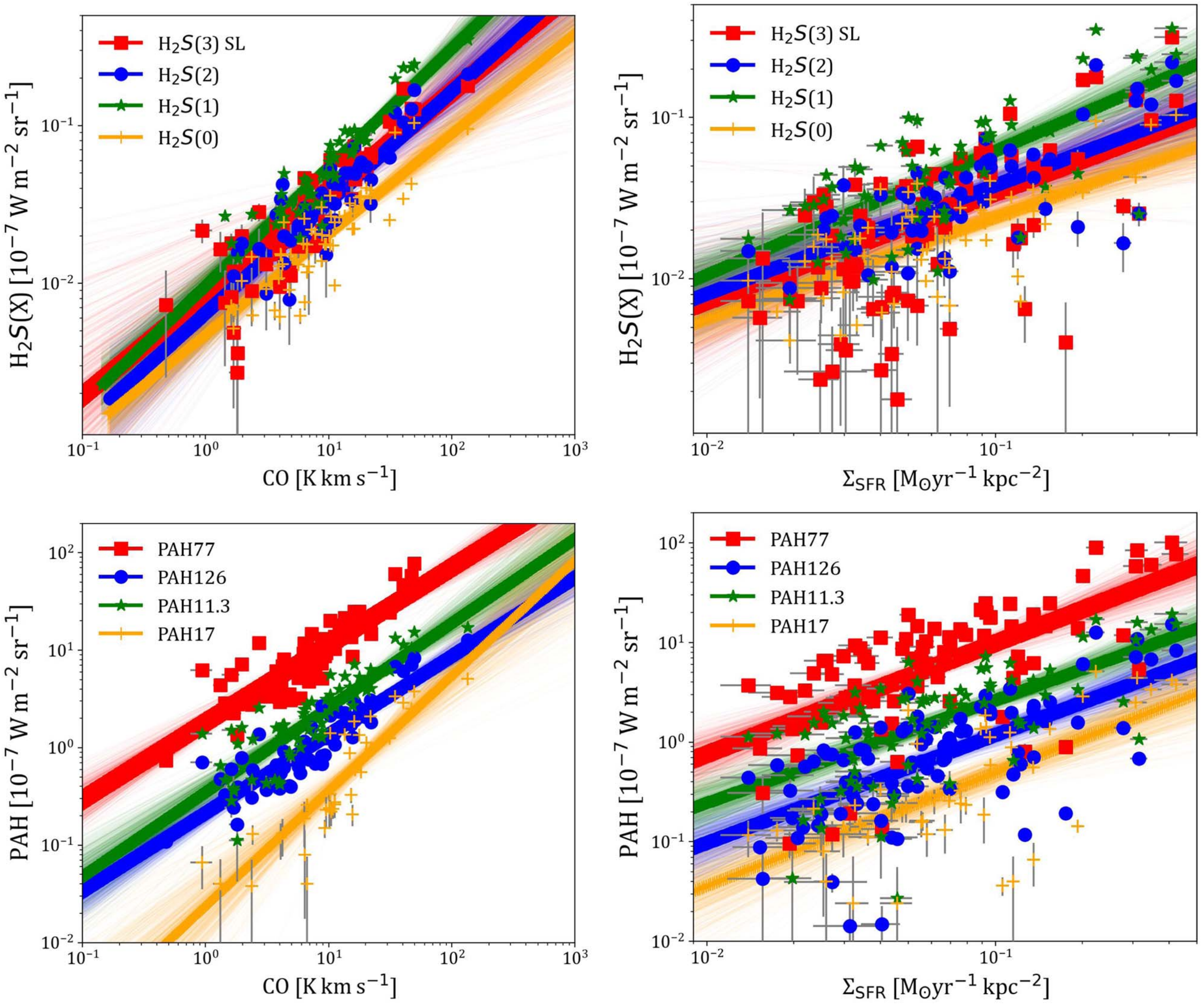

**Figure 8.** First four $H_2$ rotational lines (top) and four strongest PAH features (bottom) as a function of CO emission (left) and $\Sigma_{SFR}$ (right).

In the centers of a few of our targets there are also large deviations in CO-to-$H_2$ conversion factor due to excitation and optical depth effects (Sandstrom et al. 2013; Israel 2020). In Figure 3 we circled in red three such regions for which the CO-to-$H_2$ conversion factor is significantly different from the Milky Way value; these regions are in NGC 3184, NGC 3351, and NGC 6946, which have $\alpha_{CO}$ lower than the Milky Way value by 0.39, 0.79, and 1.04 dex, respectively (determined using dust as a tracer; Sandstrom et al. 2013). Nevertheless, the three points are not outliers in the $H_2$:CO correlation despite their different conversion factor compared to the other points. If the measured CO-to-$H_2$ conversion factors were applied, we would expect the red points to shift to lower $H_2$ mass than the others of similar CO brightness, likely increasing the scatter compared to the correlation with CO. The fact that these two points are not significant outliers in the trend between CO and $H_2$ rotational lines may suggest that the "$H_2$ rotational-to-$H_2$" conversion factor differs in those regions in the same way that CO-to-$H_2$ does. This could result from changes in the temperature distribution of the $H_2$ (either a change in the warm $H_2$ fraction or a change in the temperature distribution) coupled with changes in the temperature distribution and/or optical depth of the CO. If the $\alpha_{CO}$ changes are due to higher gas temperature, we might expect to observe warmer $H_2$ rotational temperatures in these regions, traced by $H_2$ S(3)/$H_2$ S(1). Togi & Smith (2016) also find that changes in the warm $H_2$ fraction and $H_2$ temperature may be correlated. With the three measurements currently available, we do not see clear differences in $H_2$ S(3)/$H_2$ S(1). Future studies of $H_2$ rotational ladders with JWST-MIRI in regions where $\alpha_{CO}$ varies may provide constraints on what processes can alter the CO-to-$H_2$ conversion factor.

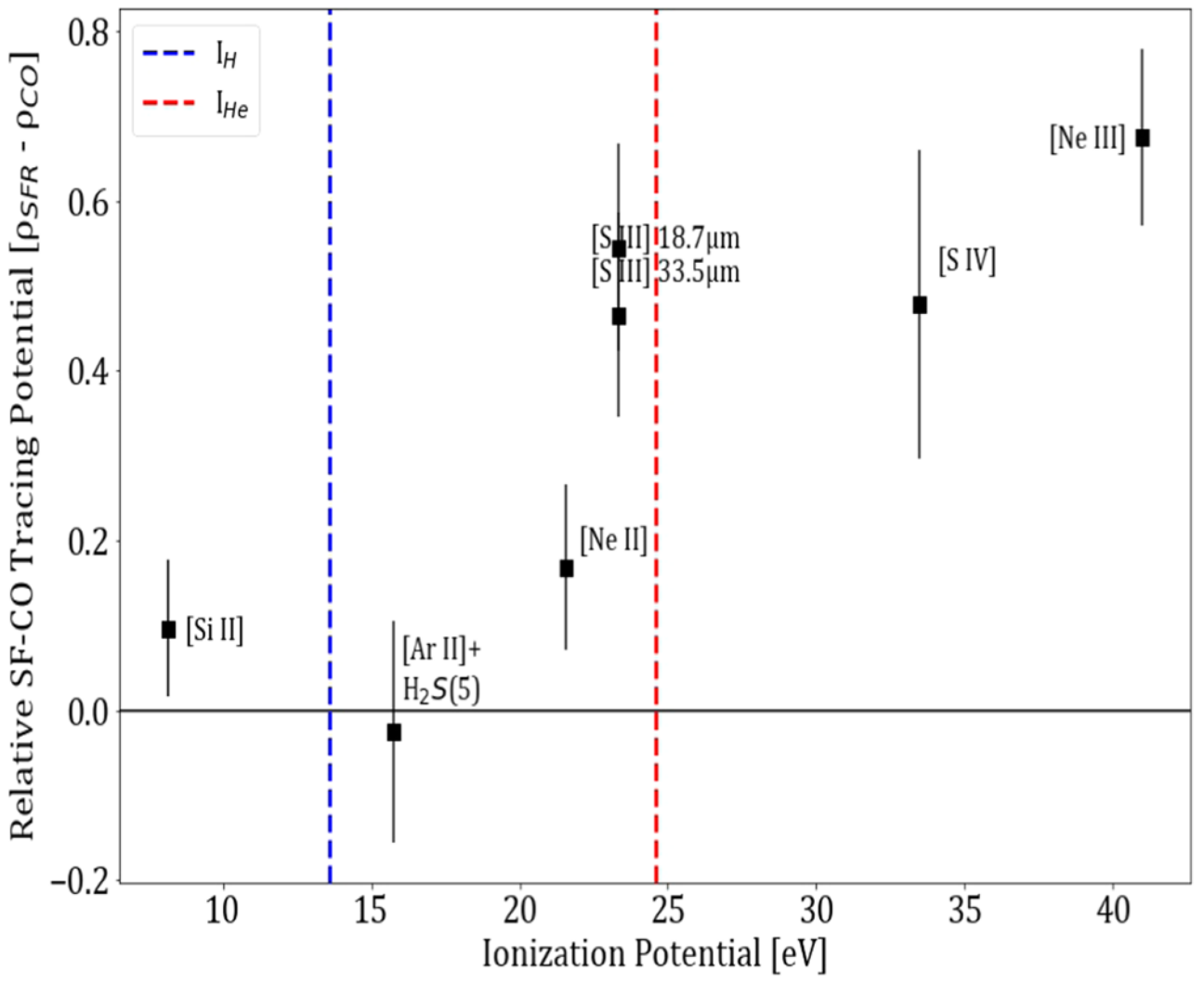

**Figure 9.** SF-CO tracing potential as a function of ionization potential for MIR fine structure lines. Vertical lines indicate the first ionization energy of hydrogen (blue) and helium (red).

### *4.3. Tracing Recent SF with Fine Structure Lines and MIR Continuum*

The only SF-dominant tracers of Figure 6 (right) in MIR spectra are the fine structure lines from ions. Most of these have ionization potentials greater than that of hydrogen. The tight correlation of these ionized gas lines and our ionizing-photon-based reference SFR tracer is expected. The lack of correlation with CO emission also implies a spatial decorrelation between lines from the most highly ionized elements and cold gas. This is especially clear in Figure 4 for the case of 15.6 $\mu$m [Ne III]. Despite being a component of our reference $\Sigma_{\rm SFR}$ tracer, we find the [Ne III]:CO correlation is significantly weaker than the K-S relation. Doubly ionizing neon requires photon energies of at least 41 eV, so the lack of correlation with cold molecular gas and strong correlation with massive SF is expected for such species.

This correlation is also exaggerated by the combination of spatial and intensity dependencies: the [Ne III] emission is not only spatially uncorrelated with CO in our smallest regions, but their intensities are driven by uncorrelated mechanisms. The [Ne III] line strength depends directly on the spectrum of ionizing radiation >41 eV, the electron density, and the abundance of $Ne^{2+}$ ions relative to hydrogen. The CO line strength depends on many factors, only one of which is an indirect dependence on the UV field intensity through PE heating of the gas by PAHs and dust.

This spatial separation between ionized and molecular gas is feasible since the median physical area of our apertures is about 0.4 $\rm kpc^2$, corresponding to a scale length of about 600 pc. Schinnerer et al. (2019) found that in maps of nearby spiral galaxies with resolution smoothed to this scale, the molecular and ionized regions (as traced by CO and H$\alpha$ emission, respectively) overlap in about 50% of pixels across the galaxy. Since the SINGS apertures were placed based on peaks of H$\alpha$ emission (Kennicutt & Armus 2003), we expect that in at least half of each region with area <0.4 $\rm kpc^2$ the $\Sigma_{\rm SFR}$ is resolved distinctly from the CO and the two are spatially uncorrelated.

In Figure 9 we show the difference between SF and CO correlation coefficients $\rho_{\rm SFR}-\rho_{\rm CO}$ for the MIR fine structure lines as a function of their ionization potential. From this, we find a trend of increasing SF dominance with increasing ionization potential. All lines are SF dominant by at least 1$\sigma$, except the blended $H_2$ S(5) and [Ar II] 6.9 $\mu$m line. $Si^+$ has a lower ionization energy than hydrogen so it can exist and emit in ionized and neutral regions. This implies [Si II] emission is SF dominant because the intensity is correlated and some of the emission may be spatially coincident with the H II region.

#### *4.3.1. MIR [S III] Lines as a Single-feature SF Tracer*

In Section 3 we showed the best correlation with the $\Sigma$Ne and Z tracer of recent SF is the [S III] fine structure line at 18.7 and/or 33.5 $\mu$m. These strong correlations with little metallicity dependence show that the MIR [S III] lines at 18.7 and/or 33.5 $\mu$m can be used to trace $\Sigma_{\rm SFR}$ as accurately as the more complex $\Sigma$Ne and Z tracer used here as our reference. The power-law slope of each [S III] fit versus $\Sigma_{\rm SFR}$ is about 1$\sigma$ greater than linear ($\sim 1.1 \pm 0.06$), indicating the emission from $S^{2+}$ is responding linearly to the radiation associated with recent SF. For the range of metallicities in our sample of H II regions in nearby galaxies, the [S III] line intensity shows no significant metallicity dependence in the scatter of its correlation with $\Sigma_{\rm SFR}$. This implies that the [S III] line intensity is a strong tracer of recent SF regardless of variations in radiation field hardness or the relative abundance of O/H, which is very well correlated with S/H abundance (Berg et al. 2020).

In our observations, fine structure lines from $S^{2+}$ have no metallicity dependence as a function of $\Sigma_{\rm SFR}$ unlike the other lines from similar ions studied here, in particular $Ne^+$ and $Ne^{2+}$. The majority of emission from doubly ionized sulfur originates from regions where photon energies are 22.3–34.8 eV (Berg et al. 2020). However, the situation is similar for singly ionized neon (22–41 eV), but we observe a strong metallicity dependence in the 12.8 $\mu$m [Ne II] correlation with SF and no significant metallicity dependence in the [S III] correlation. The next ionization state of both Ne and S also spans a similar range in excitation energy, but $Ne^{2+}$ spans a wider range from 41–64 eV while the $S^{3+}$ range is only from 35–47 eV. The other fine structure lines in our spectra, [Ar II] 6.9 $\mu$m and [Si II] 34.8 $\mu$m also have metallicity-dependent scatter as a function of SFR, in the same way as [Ne II] where low-metallicity regions emit less [Ne II] 12.8 $\mu$m, [Ar II], and [Si II] than expected based on their SFR.

Despite the similarities in the ranges of excitation energies, i.e., for $Ne^+$ and $Ne^{2+}$ and $S^{2+}$ and $S^{3+}$, we find combining [S IV] with [S III] emission results in a negligible improvement to the already tight correlation between [S III] and $\Sigma_{\rm SFR}$. As found in Whitcomb et al. (2020), we again observe the scatter in the [Ne III] correlation with SF has the opposite metallicity dependence as the [Ne II] correlation, and their sum has very little metallicity dependence. A close analog to this for sulfur would be [S IV] and [S III], but we find that only [S III] is necessary to trace SF with negligible metallicity dependence, and including [S IV] has no significant effect on the [S III] correlation with SF. The [S IV] correlation with SF is similar to that of [Ne III], with the same metallicity dependence in the scatter but even more exaggerated in the case of [S IV].

The strong correlation between SFR and [S III] appears to result from the dominance of the $S^{2+}$ ion in all H II region conditions in our sample. Previous studies have found the relative abundance of $Ne^{2+}$ to $Ne^+$ ions differs from $S^{3+}$ to $S^{2+}$ in H II regions (Rubin et al. 2008). The abundance of $S^{3+}$ never exceeds the abundance of $S^{2+}$ in H II regions in normal galaxies, while $Ne^{2+}$ frequently outnumbers $Ne^{+}$. This implies the SF correlation with MIR [S III] line emission is strong without a metallicity correction because $S^{2+}$ is the dominant form of sulfur in the star-forming regions studied here. In our previous work (Whitcomb et al. 2020) we found that the SF tracer using the sum of 15.6 $\mu$m [Ne III] and 12.8 $\mu$m [Ne II] was effective because these are the dominant forms of neon in the SINGS H II regions. However, in the case of $Ne^{2+}$ and $Ne^{+}$, both the emission line ratio and the abundance ratio vary with metallicity and radiation field intensity, so both lines are required to trace SF without residual metallicity dependence. For $S^{3+}$ and $S^{2+}$, the emission line ratio has a strong correlation with metallicity and the [Ne III] to [Ne II] line ratio, but the abundance of $S^{3+}$ to $S^{2+}$ rarely exceeds even 10% in local H II regions (Rubin et al. 2008).

A single-line, attenuation and metallicity insensitive tracer of recent SF using the [S III] 18.7 $\mu$m line is ideal for JWST MIR observations. JWST can detect this line out to redshift $z \sim 0.5$. The high sensitivity of JWST will allow studies of recent star formation rates in high-redshift galaxies of many different metallicities with the [S III] 18.7 $\mu$m line alone. Future work will be necessary to determine if the [S III]:SFR correlation begins to show significant scatter as metallicities drop below the range studied in this paper.

#### 4.3.2. Trends in IR Continuum Correlations with SF and CO

In Figure 6 (left) we found a trend as a function of wavelength where the only SF-dominant photometric bands are those between 20 and 70 $\mu$m. Above and below this range we find IR photometry traces cold molecular gas better than it traces recent star formation, with exception of the WISE3 12 $\mu$m and the PACS 100 $\mu$m bands, which are within $1\sigma$ of tracing CO and SF equally well. This trend is nearly identical to that found by Law et al. (2021; see their Figure 10) where the fraction of emission caused by young stars is modeled as a function of wavelength for the same photometric bands considered here. This strongly supports our interpretation of SF-dominant IR emission as tracers of SF, while CO-dominant IR emission traces the overall dust content.

The emission in the 20–70 $\mu$m range is dominated by high-temperature dust, so the stronger association with H II regions than cold gas we observe is consistent with findings by previous studies (e.g., Draine & Li 2007; Bendo et al. 2008; Galametz et al. 2013; Leroy et al. 2013). The peak wavelength of this thermal dust continuum is known to shift across approximately this same range from 20–70 $\mu$m or longer as the average dust temperature varies. Depending on its peak location, the fraction of thermal dust continuum captured in the adjacent WISE3 12 $\mu$m and PACS 100 $\mu$m bands will vary between regions in our sample. The portion of each of these bands due to hot dust continuum may be contributing to their strong correlation with recent star formation.

This interpretation is supported by our synthetic photometry results that show the WISE3 band is an approximately equal mix of continuum and PAH emission. As shown in Figure 2 (right), the WISE3 band measures about 55% continuum and about 40% PAH emission. In Figure 6 (left) we show the difference in SF and CO correlations for the continuum fraction of the WISE3 band separate from its PAH portion. We find the continuum portion is as SF dominant as the PAH portion is CO dominant, so the total band is about equally well correlated with SF and CO. The continuum emission below 20 $\mu$m in the SF regions considered here is dominated by the smallest and hottest dust grains, larger than PAHs and smaller than the average dust in the size distribution of carbon-based grains (Madden et al. 2006; Draine & Li 2007). These results indicate that at 12 $\mu$m, PAH emission behaves distinctly from the hot dust continuum even at the same wavelength. This could be explained by a picture where PAHs are destroyed in H II regions and the surviving dust is at very high temperatures.

We find the PAH contribution to IRAC4 8 $\mu$m is often over 80% and that from continuum about 15%. In Figure 6 (left) we find the total IRAC4 band is CO dominant, and the PAH portion is CO dominant while the continuum portion traces both CO and SF equally. The continuum at 8 $\mu$m is not SF dominant while the continuum at 12 $\mu$m is, but the uncertainty for the 8 $\mu$m continuum is large and the continuum represents a small fraction of the total emission at 8 $\mu$m.

In previous and recent studies, (e.g., Jiang et al. 2015; Gao et al. 2019, 2022; Chown et al. 2021; Leroy et al. 2023), WISE3 12 $\mu$m emission is found to trace CO emission better than H$\alpha$ traces CO at sub-kiloparsec scales. Our synthetic WISE3 results show the hot dust continuum contribution is greater than the PAH contribution in star-forming regions. In regions of galaxies with little SF, the hot dust likely contributes less to WISE3 relative to PAH emission. PAHs can be excited stochastically in diffuse and cooler gas but will be destroyed by high-energy photons near recent SF. Dust continuum at 12 $\mu$m has a stochastically excited component as well, but it also has a component that increases with the temperature of all the dust in thermal equilibrium. Near regions of recent SF, this thermal component will be significant while PAHs are destroyed. But far from these regions, the thermal component of 12 $\mu$m continuum will be small and PAHs will survive. Therefore, we expect the relative PAH:continuum fraction of the WISE3 band will increase compared to our results in regions far from SF if PAH emission increases and continuum emission remains the same or decreases.

### 4.4. Diffuse Component of PAH Emission and FIR Continuum

All PAH features are CO dominant but still correlated with SF ($\rho_{SFR}$ ~0.65), so PAH emission is associated with cold gas emission and associated with recent SF through the need for UV excitation. But there is also a significant amount of UV and optical radiation that can excite PAHs from the more numerous stars that are not recently formed O stars (Bendo et al. 2008; Crocker et al. 2013). In regions where the diffuse radiation field is stronger than those from SF, the excitation of the PAH vibrational emission will not be due to recent SF (Draine et al. 2007; Ingalls et al. 2011). In this case, the majority of PAH excitation will be due to stars less massive than O stars, implying the PAHs are not primarily tracing the most recent generation of SF but the SF averaged over long timescales.

Our investigation into the correlations between SF, CO, and the various portions of IR continuum, summarized in Figure 6 (left) has implications for the use of FIR wavelengths as SF tracers. We found that dust continuum beyond about 100 $\mu$m better traces CO emission than SF. Since dust content increases along with cold gas content, this result indicates the FIR emission is more sensitive to changes in the dust content than to the prevalence of recent SF events.

### 4.5. Implications for High-redshift IR Studies

The H II regions in this study are in galaxies within ∼20 Mpc. Galaxies at higher redshift are lower metallicity on average, so we hypothesize what the MIR correlations may look like for these high-redshift sources. Since we have few low-metallicity regions (only 12 between $\frac{1}{5}$ and $\frac{4}{5}$ $Z_{\odot}$), we only make inferences based on the presence or absence of a metallicity trend in the scatter of each correlation.

The [S III] 18.7 and 33.5 $\mu$m lines have no significant metallicity dependence in the scatter of their correlation with $\Sigma_{\rm SFR}$. The SF tracer based on either of these lines is insensitive to changes in metallicity in our sample. Other MIR tracers of SF, such as PAH or 15.6 $\mu$m [Ne III] and 12.8 $\mu$m [Ne II] emission, require explicit metallicity corrections to reproduce $\Sigma_{\rm SFR}$ values as accurate as those derived from [S III] emission alone.

The scatter in PAH emission correlations with CO is negligible, but the scatter in PAH:SF correlations has a significant metallicity dependence. Since the PAH:SF correlation fails at the low metallicities of our data set but the PAH:CO correlation does not, it is likely that PAH emission in high-redshift galaxies better traces CO than it traces SF.

The case is similar for $H_2$ rotational lines, but the $H_2$:SF scatter is less significant than in PAH:SF. This again implies that in high-redshift galaxies, the $H_2$ rotational lines will be better tracers of CO than SF, but the distinction will be less pronounced than for PAH emission at high redshift.

## 5. Conclusions

We investigate the relative ability of MIR emission features and IR photometry to trace SFRs and molecular gas content. Our data set of Spitzer-IRS spectra covering 5–35 $\mu$m from SINGS focuses on star-forming regions specifically. We compare correlation coefficients, slopes, and residual scatter for each feature and several IR continuum bands with $\Sigma_{\rm SFR}$ and with CO emission (as a proxy for $\Sigma_{\rm H2}$). We take advantage of the breakdown in the $\Sigma_{\rm SFR}$–$\Sigma_{\rm H2}$ relation at small scales in order to find spectral features that trace $\Sigma_{\rm SFR}$ distinctly from those that trace $\Sigma_{\rm H2}$.

From this investigation we observe the following:

1. The MIR features that best correlate with CO $J = (2-1)$ are the $H_2$ S(0) and S(1) rotational lines of molecular hydrogen, and all the major PAH features. We attribute the strong mutual correlation between PAH, $H_2$, and CO emission to a combination of spatial and intensity correlations (i.e., each arises in molecular gas and each is excited by UV photons, either directly or indirectly). The PAH emission bands are CO dominant, and all $H_2$ rotational lines and PAH features are better correlated with $\Sigma_{\rm SFR}$ than CO is.
2. The PAH feature ratios vary as a function of CO emission, but not as a function of $\Sigma_{\rm SFR}$. We find the longer wavelength PAH features are brighter relative to short-wavelength features, e.g., 17–7.7 $\mu$m, with increasing CO brightness. This may be evidence of PAH growth and/or neutralization in molecular gas, which has been proposed in several previous studies.
3. The scale dependence of the K-S star formation relation causes all correlations with SF and CO to be strong in the larger regions of our SINGS H II region sample. However, at scales $\lesssim$0.4 kpc$^2$ we find SF and CO are uncorrelated and the different IR emission features show a better correlation with either SF or CO.
4. We calibrate a single-line tracer of recent star formation based on the [S III] 18.7 or 33.5 $\mu$m line intensity. This tracer has a negligible metallicity dependence in our sample and predicts SFR equivalent to the $\Sigma$Ne and Z tracer which explicitly includes a metallicity correction.
5. We observe a trend as a function of wavelength in the correlation of IR photometry with SF and CO, such that the infrared continuum between about 20 and 70 $\mu$m is strongly SF dominant, but further from this range, the emission is increasingly CO dominant. From this trend as shown in Figure 6 (left), we hypothesize that the continuum between 20 and 70 $\mu$m may be as strong an SF tracer as MIPS 24 $\mu$m and PACS 70 $\mu$m. The trend is consistent with that of the fraction of IR emission caused by young stars as a function of wavelength, indicating our method is distinguishing tracers of recent star formation from tracers of the dust content and interstellar radiation field at sub-kiloparsec scales.
6. The WISE3 12 $\mu$m band is less than 50% PAH emission (median ∼40%), with the dominant portion being dust continuum emission from hot, small dust grains. We find the continuum at 12 $\mu$m better traces SF while the PAH features better trace CO. The IRAC4 8 $\mu$m band is dominated by the 7.7 $\mu$m PAH complex with a median composition of 78%:16%:3%:4% PAH:Continuum:Starlight:Lines.

In the SINGS H II regions we find that all dust-based IR emission features from 3.4–250 $\mu$m are primarily tracing the amount of dust with various differences in their response to the local heating. Emission lines from ionized gas are the strongest MIR tracers of $\Sigma_{\rm SFR}$ and the weakest MIR tracers of CO in our sample. Using dust-based $\Sigma_{\rm SFR}$ and $\Sigma_{\rm H2}$ tracers to study the star formation relation in galaxies will give an artificially enhanced correlation because they all trace both SF and $H_2$ to varying degrees.

We thank the anonymous referee for useful comments that improved the final manuscript. This work uses observations made with the Spitzer Space Telescope, which is operated by the Jet Propulsion Laboratory, California Institute of Technology, under a contract with NASA. This research has made use of NASA's Astrophysics Data System. This research made use of Astropy, a community-developed core Python package for Astronomy (Astropy Collaboration et al. 2018, 2013), synphot (STScI Development Team 2018), SciPy (Virtanen et al. 2020), NumPy (Harris et al. 2020) and matplotlib, a Python library for publication quality graphics (Hunter 2007). The authors acknowledge the efforts of the SINGS Team (2020); KINGFISH Team (2020), z0MGS (Sandstrom 2019), and PHANGS teams in assembling public data sets used in this work. CMW thanks G. Donnelly for creating an acronym for our spectral fitting algorithm. C.M.W., J.D.S., and K.S. acknowledge funding support from NASA ADAP grant 80NSSC21K0851. A.K.L. gratefully acknowledges support by grants 1653300 and 2205628 from the National Science Foundation, by award JWST-GO-02107.009-A, and by a Humboldt Research Award from the Alexander von Humboldt Foundation.

This paper makes use of the following ALMA data: ADS/JAO.ALMA#2012.1.00650.S, ADS/JAO.ALMA#2015.1.00782.S, ADS/JAO.ALMA#2015.1.00925.S, ADS/JAO.ALMA#2015.1.00956.S, ADS/JAO.ALMA#2017.1.00392.S, ADS/JAO.ALMA#2017.1.00886.L, ADS/JAO.ALMA#

2018.1.01651.S, ALMA is a partnership of ESO (representing its member states), NSF (USA), and NINS (Japan), together with NRC (Canada), NSC and ASIAA (Taiwan), and KASI (Republic of Korea), in cooperation with the Republic of Chile. The Joint ALMA Observatory is operated by ESO, AUI/NRAO, and NAOJ. The National Radio Astronomy Observatory is a facility of the National Science Foundation operated under cooperative agreement by Associated Universities, Inc.

## Appendix

### *A.1. Description of Region Data Table*

Table 5 shows the general form and content of the machine-readable table of data used in this work. This includes the coordinates of the vertices of the 154 apertures that define the overlap between SINGS SL, SH, and LH coverage, as well as all data derived for each: the aperture angular and physical area, the SFR surface density, integrated intensities of CO $J = (2 - 1)$ and all IR photometry and spectroscopic features described in Section 2.

### *A.2. Spectral Fitting Comparisons*

Figure 10 shows a comparison between PAHFIT and FARAMIR integrated intensities (see Section 2.7) for four emission features in the overlapping wavelengths of the SL and SH suborders of Spitzer-IRS (∼10 to 15 $\mu$m). PAHFIT is able to fit dimmer PAH features than FARAMIR, but the agreement is strong at higher intensities.

**Table 5**
Description of Region Data Table

| | Feature | Description |
|---|---|---|
| 1 | PAH62 SL | 6.2 $\mu$m PAH SL integrated intensity and uncertainty |
| 2 | PAH77 SL | 7.7 $\mu$m PAH SL integrated intensity and uncertainty |
| 3 | PAH83 SL | 8.3 $\mu$m PAH SL integrated intensity and uncertainty |
| 4 | PAH86 SL | 8.6 $\mu$m PAH SL integrated intensity and uncertainty |
| 5 | PAH113 SL | 11.3 $\mu$m PAH SL integrated intensity and uncertainty |
| 6 | PAH120 SL | 12.0 $\mu$m PAH SL integrated intensity and uncertainty |
| 7 | PAH126 SL | 12.6 $\mu$m PAH SL integrated intensity and uncertainty |
| 8 | PAH136 SL | 13.6 $\mu$m PAH SL integrated intensity and uncertainty |
| 9 | PAH142 SL | 14.2 $\mu$m PAH SL integrated intensity and uncertainty |
| 10 | [Ar II] | blended $H_2S(5)$ and [Ar II] integrated intensities and uncertainty |
| 11 | $H_2S(3)$ SL | $H_2S(3)$ integrated intensity and uncertainty |
| 12 | $H_2S(2)$ SL | $H_2S(2)$ SL integrated intensity and uncertainty |
| 13 | [Ne II] SL | [Ne II] SL integrated intensity and uncertainty |
| 14 | IRAC 3.6 $\mu$m | IRAC 3.6 $\mu$m integrated intensity and uncertainty |
| 15 | IRAC 4.5 $\mu$m | IRAC 4.5 $\mu$m integrated intensity and uncertainty |
| 16 | IRAC 5.8 $\mu$m | IRAC 5.8 $\mu$m integrated intensity and uncertainty |
| 17 | IRAC 8 $\mu$m | IRAC 8 $\mu$m integrated intensity and uncertainty |
| 18 | WISE 3.4 $\mu$m | WISE 3.4 $\mu$m integrated intensity and uncertainty |
| 19 | WISE 4.6 $\mu$m | WISE 4.6 $\mu$m integrated intensity and uncertainty |
| 20 | WISE 12 $\mu$m | WISE 12 $\mu$m integrated intensity and uncertainty |
| 21 | WISE 22 $\mu$m | WISE 22 $\mu$m integrated intensity and uncertainty |
| 22 | MIPS 24 $\mu$m | MIPS 24 $\mu$m integrated intensity and uncertainty |
| 23 | PACS 70 $\mu$m | PACS 70 $\mu$m integrated intensity and uncertainty |
| 24 | PACS 100 $\mu$m | PACS 100 $\mu$m integrated intensity and uncertainty |
| 25 | PACS 160 $\mu$m | PACS 160 $\mu$m integrated intensity and uncertainty |
| 26 | SPIRE 250 $\mu$m | SPIRE 250 $\mu$m integrated intensity and uncertainty |
| 27 | CO | CO $J = (2 - 1)$ integrated intensity and uncertainty |
| 28 | [S IV] | [S IV] integrated intensity and uncertainty |
| 29 | PAH 11.0 $\mu$m | 11.0 $\mu$m PAH SH integrated intensity and uncertainty |
| 30 | PAH 11.2 $\mu$m | 11.2 $\mu$m PAH SH integrated intensity and uncertainty |
| 31 | PAH 11.25 $\mu$m | 11.25 $\mu$m PAH SH integrated intensity and uncertainty |
| 32 | PAH 11.4 $\mu$m | 11.4 $\mu$m PAH SH integrated intensity and uncertainty |
| 33 | PAH 12.0 $\mu$m | 12.0 $\mu$m PAH SH integrated intensity and uncertainty |
| 34 | $H_2S(2)$ | $H_2S(2)$ SH integrated intensity and uncertainty |
| 35 | Hu-alpha | Humphreys-$\alpha$ integrated intensity and uncertainty |
| 36 | PAH 12.6 $\mu$m | 12.6 $\mu$m PAH SH integrated intensity and uncertainty |
| 37 | PAH 12.7 $\mu$m | 12.7 $\mu$m PAH SH integrated intensity and uncertainty |
| 38 | [Ne II] | [Ne II] SH integrated intensity and uncertainty |
| 39 | PAH 13.5 $\mu$m | 13.5 $\mu$m PAH SH integrated intensity and uncertainty |
| 40 | PAH 14.2 $\mu$m | 14.2 $\mu$m PAH SH integrated intensity and uncertainty |
| 41 | [Ne V] | [Ne V] integrated intensity and uncertainty |
| 42 | [Ne III] | [Ne III] integrated intensity and uncertainty |
| 43 | PAH 15.9 um | 15.9 $\mu$m PAH SH integrated intensity and uncertainty |
| 44 | PAH 16.5 um | 16.5 $\mu$m PAH SH integrated intensity and uncertainty |
| 45 | $H_2S(1)$ | $H_2S(1)$ integrated intensity and uncertainty |
| 46 | PAH 17.0 $\mu$m | 17.0 $\mu$m PAH SH integrated intensity and uncertainty |

**Table 5**
(Continued)

| | Feature | Description |
|---|---|---|
| 47 | PAH 17.4 $\mu$m | 17.4 $\mu$m PAH SH integrated intensity and uncertainty |
| 48 | PAH 17.9 $\mu$m | 17.9 $\mu$m PAH SH integrated intensity and uncertainty |
| 49 | [S III] SH | 18.7 $\mu$m [S III] integrated intensity and uncertainty |
| 50 | PAH 18.9 $\mu$m | 18.9 $\mu$m PAH SH integrated intensity and uncertainty |
| 51 | $H_2S(0)$ | $H_2S(0)$ integrated intensity and uncertainty |
| 52 | [S III] | 33.5 $\mu$m [S III] integrated intensity and uncertainty |
| 53 | [Si II] | [Si II] integrated intensity and uncertainty |
| 54 | PAHs <12 $\mu$m | Total PAH <12.0 $\mu$m integrated intensity and uncertainty |
| 55 | PAHs >12 $\mu$m | Total PAH >12.0 $\mu$m integrated intensity and uncertainty |
| 56 | Total 11.3 $\mu$m | 11.3 $\mu$m PAH complex SH integrated intensity and uncertainty |
| 57 | Total 12.6 $\mu$m | 12.6 $\mu$m PAH complex SH integrated intensity and uncertainty |
| 58 | Total 17 $\mu$m | 17 $\mu$m PAH complex SH integrated intensity and uncertainty |
| 59 | Total neon | [Ne II]+[Ne III] integrated intensity and uncertainty |
| 60 | Total [S III] | 18.7 $\mu$m [S III]+ 33.5 $\mu$m [S III] integrated intensity and uncertainty |
| 61 | [S III]+[S IV] | 18.7 $\mu$m [S III]+[S IV] integrated intensity and uncertainty |
| 62 | Total $H_2S$ | $H_2S(0)+S(1)+S(2)+S(3)$ integrated intensity and uncertainty |
| 63 | $H_2S(0+1+2)$ | $H_2S(0)+S(1)+S(2)$ integrated intensity and uncertainty |
| 64 | TIR | Total infrared integrated intensity and uncertainty |
| 65 | SynthWISE: cont. | Continuum portion of WISE 12 $\mu$m and uncertainty |
| 66 | SynthWISE: PAH | PAH portion of WISE 12 $\mu$m and uncertainty |
| 67 | SynthIRAC: cont. | Continuum portion of IRAC 8 $\mu$m and uncertainty |
| 68 | SynthIRAC: PAH | PAH portion of IRAC 8 $\mu$m and uncertainty |
| 69 | $H_2S(1)$/PAH 7.7 $\mu$m | $H_2S(1)$ to 7.7 $\mu$m PAH ratio and uncertainty |
| 70 | [Ne III]/[Ne II] | [Ne III] to [Ne II] ratio and uncertainty |
| 71 | [S III] 19/34 $\mu$m | 18.7 $\mu$m [S III] to 33.5 $\mu$m [S III] ratio and uncertainty |
| 72 | $12 + \log_{10}$[O/H] | KK04 metallicity and uncertainty |
| 73 | SFR | SFR surface density and uncertainty |
| 74 | Distance | Distance to host galaxy |
| 75 | Angular area | Angular area of region |
| 76 | Deprojected area | Deprojected physical area of region |
| 77 | Vertex 1 | Vertex 1 R.A. and decl. |
| 78 | Vertex 2 | Vertex 2 R.A. and decl. |
| 79 | Vertex 3 | Vertex 3 R.A. and decl. |
| 80 | Vertex 4 | Vertex 4 R.A. and decl. |
| 81 | Vertex 5 | Vertex 5 R.A. and decl. |
| 82 | Vertex 6 | Vertex 6 R.A. and decl. |
| 83 | Vertex 7 | Vertex 7 R.A. and decl. |

**Note.** The integrated intensities have units of $10^{-7}$ W m$^{-2}$ sr$^{-1}$. The SFR surface densities have units of $M_\odot$ yr$^{-1}$ kpc$^{-2}$. The distances are in units of megaparsec. The angular areas are in units of square arcseconds. The deprojected areas are in units of square kiloparsec. The vertex locations are in units of decimal degrees.

(This table is available in its entirety in machine-readable form.)

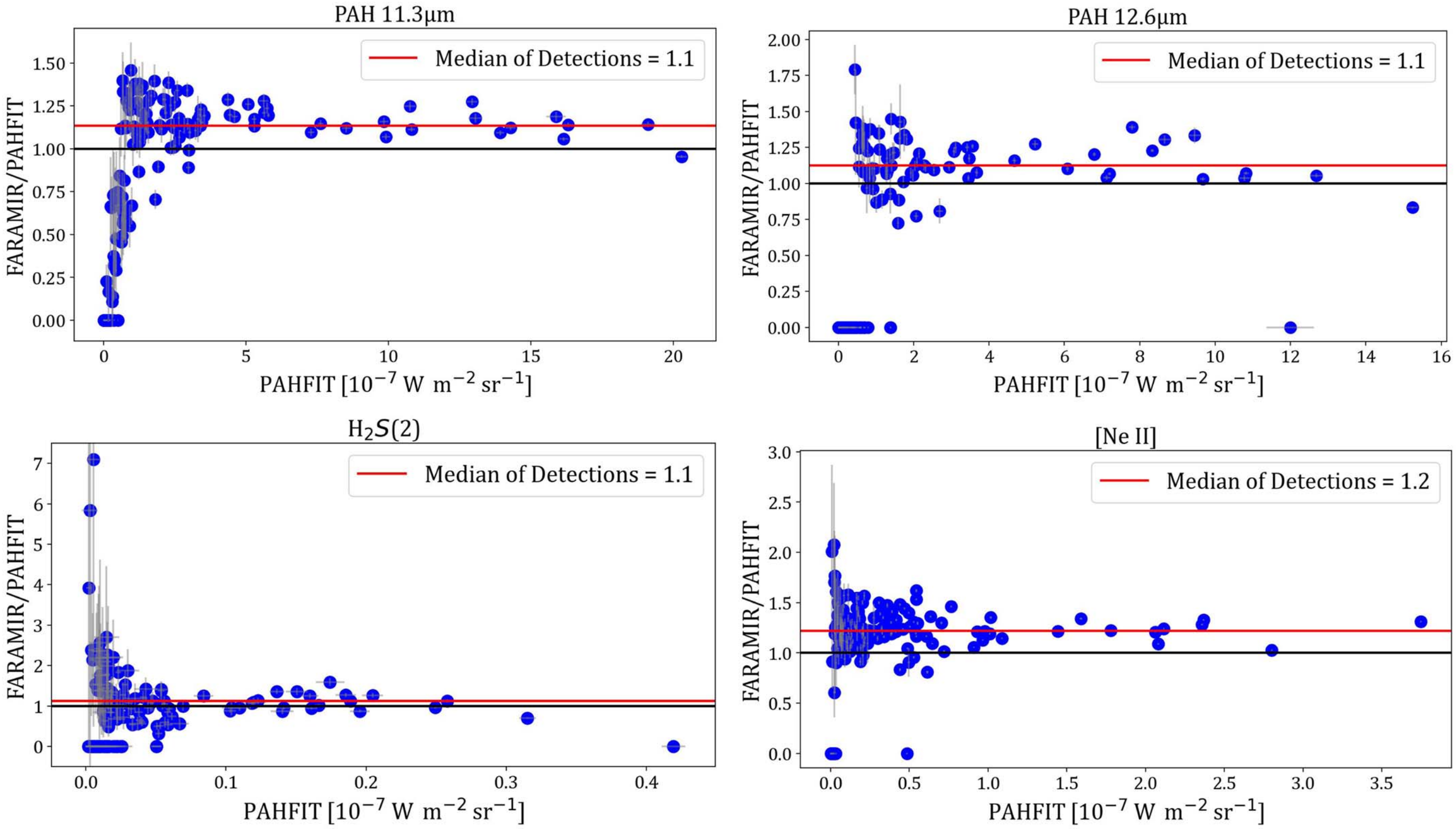

**Figure 10.** Comparison between results obtained by the fitting technique described in Section 2.7 (FARAMIR) on SH spectra relative to those returned by PAHFIT on corresponding SL spectra for the 11.3 $\mu$m PAH complex (top left), the 12.6 $\mu$m PAH complex (top right), the $H_2$ S(2) rotational line (bottom left), and the 12.8 $\mu$m [Ne II] fine structure line (bottom right). The median of the FARAMIR to PAHFIT ratios is indicated by the red line in each panel, non-detections are excluded.

## ORCID iDs

C. M. Whitcomb https://orcid.org/0000-0003-2093-4452
K. Sandstrom https://orcid.org/0000-0002-4378-8534
A. Leroy https://orcid.org/0000-0002-2545-1700
J.-D. T. Smith https://orcid.org/0000-0003-1545-5078